\documentclass{JHEP3} 

\usepackage{epsfig,multicol}
\usepackage{amsmath, amssymb}
\usepackage{epic}
\usepackage{concmath, palatino}

\newcommand\fverb{\setbox\pippobox=\hbox\bgroup\verb}
\newcommand\fverbdo{\egroup\medskip\noindent%
            \fbox{\unhbox\pippobox}\ }
\newcommand\fverbit{\egroup\item[\fbox{\unhbox\pippobox}]}
\newbox\pippobox

\newcommand{\be}{\begin{equation}}
\newcommand{\ee}{\end{equation}}
\newcommand{\ba}{\begin{eqnarray}}
\newcommand{\ea}{\end{eqnarray}}

\newcommand{\refeq}[1]{Eq.~(\ref{eq:#1})}

\newcommand{\ads}{AdS_5\times S^5}
\newcommand{\tran}{transcendentality\ }

\newtheorem{theorem}{Theorem}[section]

\title{Reciprocity of gauge operators in ${\cal N}=4$ SYM}

\author{Matteo Beccaria\\
  Dipartimento di Fisica, Universita' del Salento,
  Via Arnesano, I-73100 Lecce\\
  INFN, Sezione di Lecce\\
  E-mail: \email{matteo.beccaria@le.infn.it}}

\author{Valentina Forini\\
  Humboldt-Universit\"{a}t zu Berlin, Institut f\"{u}r Physik, Newtonstra{\ss}e
  15, D-12489 Berlin\\
  E-mail: \email{forini@physik.hu-berlin.de}}

\preprint{HU-EP-08/07}

\abstract{
A recently discovered  generalized Gribov-Lipatov reciprocity holds for the anomalous dimensions of various twist operators in ${\cal N}=4$ SYM.
Here, we consider a class of scaling $\mathfrak{psu}(2,2|4)$ operators that reduce at one-loop to twist-3 maximal helicity gluonic operators.
We extract from the asymptotic long-range Bethe Ansatz a closed expression for the spin dependent anomalous dimension at four loop order and 
provide a complete proof of reciprocity. We comment about the interplay with possible, yet unknown, wrapping corrections.}


\begin{document}

\allowdisplaybreaks

\section{Introduction}
\label{Sec:Intro}

The maximally supersymmetric theory ${\cal N}=4$ SYM is dual to type II superstring on $\ads$ and plays a central role in the AdS/CFT correspondence~\cite{Maldacena:1997re}.
The existence of a strong-weak coupling duality links the integrability properties on the string side~\cite{Bena:2003wd} to 
a well-known form of {\em internal} integrability in the superconformal theory~\cite{Beisert:2004ry}. At one-loop, 
the scale dependence of renormalized composite operators
is governed in the planar limit by a local integrable super spin-chain Hamiltonian~\cite{Beisert:2003jj}. At higher loops, 
integrability persists and is described by a long-range lattice Hamiltonian whose interaction range increases with the loop order~\cite{Beisert:2005fw}. In particular, AdS/CFT duality has been crucial in prompting the higher loop proposal for the $S$-matrix of $\mathcal{N}=4$ SYM theory~\cite{Arutyunov:2004vx,Staudacher:2004tk,Park:2005ji,Beisert:2005tm,Janik:2006dc,Eden:2006rx,Hernandez:2006tk,Freyhult:2006vr,Plefka:2006ze,Beisert:2006ib,Beisert:2006ez}.

The energy levels of the integrable spin-chain compute the anomalous dimension of scaling fields in the superconformal theory, {\em i.e.}
the energies of would-be dual string states.
For a given specific operator, the calculation amounts to finding the relevant solution of a rather complicated set of Bethe Ansatz equations.
Of course, finding a closed formula for a {\em class} of operators is a more difficult task. In some applications, aimed at accurate tests of AdS/CFT duality,
one considers operators which are gauge invariant single traces with varying length. In the large length limit, the size corrections can be computed by a thermodynamical 
analysis of the Bethe Ansatz equations~\cite{Beisert:2003xu,Beisert:2003ea}. 
In exceptional cases, it is also possible to exhibit closed formulae for the anomalous dimensions at finite length~\cite{Beccaria:2007gu}.

Here, we shall be interested in the  class of so-called \emph{quasipartonic} twist operators~\cite{quasipart}. They have a basically fixed field content, 
but are constructed with an arbitrary number of covariant derivatives distributed among the fields. The twist operators are characterized by a simple control
parameter which is the total number of derivatives, simply related to the total Lorentz spin $N$. From the spin-chain point of view, they 
are associated with fixed length states, at least in the one-loop description of mixing.
The thermodynamical limit of a large number of Bethe roots is nothing but the large spin limit $N\to\infty$ and in this regime
it is possible to derive integral equations computing the roots distribution at all orders in the gauge coupling~\cite{Eden:2006rx,Beisert:2006ez,Rej:2007vm}.

Surprisingly, in some cases it is also possible to provide closed multi-loop expressions for the anomalous dimension $\gamma(N)$
of special twist operators as functions of the Lorentz spin $N$~\cite{Staudacher:2004tk,Kotikov:2007cy,Beccaria:2007cn,Beccaria:2007vh,Beccaria:2007bb,Beccaria:2007pb}.
Currently, it is not known how to derive systematically the functions $\gamma(N)$ beyond the one-loop level although some progress can be done exploiting the 
Baxter approach~\footnote{A.~V.~Kotikov, private communication.}. Recent analytical attempts are discussed in~\cite{Belitsky:2007jp,Beccaria:2007uj}.

The anomalous dimensions $\gamma(N)$ are expected to contain important information encoded in their dependence on $N$.
The physical content of this information can be extracted by exploiting known facts valid for similar twist operators arising in the QCD analysis of 
deep inelastic scattering (DIS)~\cite{Altarelli:1981ax,Martin:2008cn}.
In that context, one can consider the leading twist-2 contributions and connect the total spin $N$ to its dual, in Mellin space, which is the Bjorken variable $x$.  Two opposite regimes emerge in a natural way, each carrying its lore of approximations.  
The first is small $x\to 0$ and is captured by the BFKL equation~\cite{Lipatov:1976zz}. It can be analyzed by considering 
the Regge poles of $\gamma(N)$ analytically continued to negative (unphysical) values of the spin. A recent detailed example of such analysis is discussed in~\cite{Kotikov:2007cy}.

Here, we shall be interested in the properties of the second {\em quasi-elastic} regime which is $x\to 1$, \emph{i.e.} large $N$.
The following general features can be inferred from the large $N$ behavior of known three loops twist-2 QCD results
as well as from general results valid at higher twist~\cite{Belitsky:2003ys}
\begin{enumerate}
\item The leading large $N$ behavior of the anomalous dimensions $\gamma(N)$ is logarithmic 
\be
\gamma(N) = 2\, \Gamma(\alpha_s)\,\log\,N + {\cal O}(N^0),\qquad N\to\infty.
\ee
The function $\Gamma(\alpha_s)$ is a universal function of the coupling related to soft gluon emission~\cite{Korchemsky:1992xv,Belitsky:2003ys,Belitsky:2006en}.
It appears as a cusp anomalous dimension governing the renormalization of a light-cone Wilson loop describing soft-emission processes as quasi-classical charge motion.

\item The subleading terms in the large $N$ expansion of $\gamma(N)$ obey (three loops) hidden relations, the Moch-Vermaseren-Vogt (MVV) constraints~\cite{Moch:2004pa,Vogt:2004mw}.
Recently, they have been extended to an infinite set of  higher orders relations in the $1/N$ expansion~\cite{Basso:2006nk}. Basically, they predict that 
roughly half of the $1/N$ expansion is completely determined by the other half.
\end{enumerate}

A very promising strategy is certainly that of investigating these features in the context of planar ${\cal N}=4$ SYM, where integrability 
techniques afford a relatively painless multi-loop analysis. This approach could shed light on the otherwise elusive beautiful structures found in the 
closed expressions of twist anomalous dimensions.

>From this point of view, we can reconsider point 1. in the above list. 
It is well known that an integral equation has been derived providing the all-order weak coupling expansion of $\Gamma(\alpha_s)$~\cite{Eden:2006rx,Beisert:2006ez}. 
The calculation has been extended at strong-coupling in the explicit case of the $\mathfrak{sl}(2)$ 
sector~\cite{StrongSL2} and is amenable to 
wide generalizations~\cite{Kruczenski:2008bs}.
Thus, our attitude is that the general remark 1. is a strong check for any guessed expression $\gamma(N)$ describing a particular class of twist operators.

Concerning 2., the understanding of MVV relations is instead more intriguing and less conclusive. In the twist-2 QCD context, it is 
known that the existance of MVV relations is related with space-time reciprocity of DIS and its crossed version of $e^+e^-$ annihilation into 
hadrons (see~\cite{Dokshitzer:2007zz} for a very clear pedagogical discussion). This is a non-trivial all-order generalization of the one-loop Gribov-Lipatov (GL) 
reciprocity~\cite{Gribov:1972rt}. Positive three loops tests for QCD and for the universal twist-2 supermultiplet in ${\cal N}=4$ SYM are discussed in~\cite{Basso:2006nk,Dokshitzer:2006nm}.
Technically, reciprocity in the twist-2 case  holds for the Dokshitzer-Marchesini-Salam (DMS) evolution kernel governing simultaneously the
distribution and fragmentation functions~\cite{Dokshitzer:2005bf}. The MVV relations follow as a straightforward corollary.

The formalism of the DMS reciprocity respecting kernel can be extended to higher twists and in particular to various twist-3 sectors where closed 
formulae for the anomalous dimensions are available in ${\cal N}=4$ SYM. Remarkably, the generalized Gribov-Lipatov reciprocity works perfectly.
The first example is the relatively simple $\mathfrak{sl}(2)$ sector~\cite{Beccaria:2007bb}, where a 4 loops complete proof is available. Additional evidences of reciprocity 
for fermionic and gauge operators (both at three loops) have been later discussed in~\cite{Beccaria:2007pb,Beccaria:2007vh}.

Here, we present a complete analysis of a nested gluonic sector~\cite{Beccaria:2007pb,Belitsky:2008wj} that we study at four 
loops. Our main result is that reciprocity holds rigorously even in this case, modulo possible wrapping effects.

This paper is organized as follows: In Sec.~(\ref{sec:qcd}), as a reminder,  we recall the main QCD facts concerninig the generalized Gribov-Lipatov
reciprocity. In Sec.~(\ref{sec:n4}) we present the suitable extension to ${\cal N}=4$ SYM theory with a summary of 
known successfull reciprocity tests. 
In Sec.~(\ref{sec:gluonic}) we discuss in details the class of operators studied in this paper. The four loop 
anomalous dimension is presented in Sec.~(\ref{sec:fourloops}), and a complete proof of its reciprocity properties is 
derived in Sec.~(\ref{sec:proof}). Finally, Sec.~(\ref{sec:wrapping}) contains a discussion of the relation beween reciprocity and wrapping.
In Appendix~(\ref{app:expansions}), we collect several tests of our results related to the large $N$ expansion. 
Appendix~(\ref{app:harmonic}) is a short primer on nested harmonic sums collecting useful definitions and formulae. 

\section{Reciprocity of twist-2 anomalous dimensions in QCD}
\label{sec:qcd}

\subsection{Gribov-Lipatov reciprocity}

The scale dependence of QCD parton distribution functions in deep inelastic scattering is governed by the 
the DGLAP evolution equations~\cite{Gribov:1972rt,Altarelli:1977zs,Dokshitzer:1977sg}. The non perturbative ingredients are 
the space-like (S) splitting functions $P_S(x)$, related to the anomalous dimensions of twist-2 operators~\cite{Brock:1993sz} through 
a Mellin transformation.
Three loop results for the anomalous dimensions $\gamma_S(N)$ governing the evolution of singlet and 
non-singlet unpolarized distributions have been obtained in \cite{Moch:2004pa,Vogt:2004mw}.

The related crossed process of $e^+e^-$ annihilation into hadrons involves the non perturbative fragmentation functions. In their 
scale evolution the role of splitting functions is played by the so-called time-like (T) splitting functions $P_T(x)$, which allow to define
time-like anomalous dimensions $\gamma_T(N)$ again by a Mellin transformation.
A basic question is then: {\it What is the relation between space and time-like kernels $P_S$ and $P_T$ ?}

A first relation between $P_S(x)$ and $P_T(x)$ is the Drell-Levy-Yan relation~\cite{DLY}
\be
\mbox{Drell-Levy-Yan}:\qquad P_T(x) = -\frac{1}{x}\,P_S\left(\frac{1}{x}\right).
\ee
This is an analytic continuation from one kernel to the other which passes through the singular point $x=1$ at the border of the respective disjoint physical regions. 
It is a relation trivial at one-loop and full of subtleties at higher orders. A discussion at two loops is presented in~\cite{Stratmann:1996hn}.

A second equation has been proposed by Gribov and Lipatov~\cite{Gribov:1972rt}, that reads
\be
\mbox{Gribov-Lipatov}:\qquad P_T(x) = P_S(x) \equiv P(x).
\ee
Assuming this result and the (true) Drell-Levy-Yan relation, we get
the following \emph{reciprocity} for the common function $P(x)$ 
\be
\mbox{Gribov-Lipatov reciprocity}:\qquad P(x) = -x\,P\left(\frac{1}{x}\right).
\ee
In Mellin space~\footnote{
The Mellin transform ${\cal F}(N)$ of $f(x)$ is defined by ${\cal F}(N) = \int_0^1\,dx\,x^{N-1}\,f(x)$.
}, 
it can be shown that this means (in the sense of asymptotic expansions at large $N$) 
\be
P(N) = f(J^2), \qquad J^2 = N\,(N+1),\qquad N\to\infty.
\ee
Gribov-Lipatov reciprocity holds at one-loop, but fails at two loops~\cite{Curci:1980uw,Furmanski:1980cm}.
The explicit violation can be written as
\be
\label{eq:violation}
\frac{1}{2}\left[P_{T, qq}^{(2)}(x)-P_{S, qq}^{(2)}\right] = \int_0^1\frac{dz}{z}\left\{P_{qq}^{(1)}\left(\frac{x}{z}\right)\right\}_+\,P_{qq}^{(1)}(z)\,\log\,z.
\ee
It is {\em kinematic} in the sense that it is entirely expressed in terms of the one-loop kernel. A deep explanation for this naive observation is illustrated
in the next section.

\subsection{Reciprocity respecting evolution equations}

The evolution equations for the parton distributions or fragmentation functions $D_\sigma(x, Q^2)$ ($\sigma = S, T$) take the 
standard convolution form 
\be
\partial_\tau\,D_\sigma(x, Q^2) = \int_0^1\frac{dz}{z}\,P_\sigma(z, \alpha_s(Q^2))\,D_\sigma\left(\frac{x}{z}, Q^2\right),
\ee
where $P_\sigma$ are the space or time-like splitting functions and $\tau = \log\,Q^2$.
 Mellin transforming, this reads
\be
\partial_\tau\,D_\sigma(N, Q^2) = -\frac{1}{2}\gamma_\sigma(N, \alpha_s(Q^2))\, D_\sigma(N, Q^2),
\ee
where
\be
D_\sigma(N, Q^2) = \int_0^1\frac{dx}{x}\,x^N\,D_\sigma(x, Q^2),\qquad
\gamma_\sigma(N, Q^2) = -\frac{1}{2}\int_0^1\frac{dx}{x}\,x^N\,P_\sigma(x, \alpha_s(Q^2)).
\ee
Based on several deep physical ideas, it has been proposed to rewrite the evolution equation in a way that aims at treating the DIS and $e^+e^-$ channels
more symmetrically, in the spirit of Gribov-Lipatov reciprocity~\cite{Dokshitzer:1995ev,Dokshitzer:2005bf}.
The reciprocity respecting evolution equations  take the form 
\be
\label{eq:REE}
\partial_\tau\,D_\sigma(x, Q^2) = \int_0^1\frac{dz}{z}\,{\cal P}(z)\,D_\sigma\left(\frac{x}{z}, z^\sigma\,Q^2\right),
\ee
where $\sigma=-1, 1$ for the space and time like channels respectively. In the equation above we have not written in details the scale dependence of the coupling for reasons to be explained later.

The crucial point is that the evolution kernel ${\cal P}(z)$ is the same in both channels. 
As an immediate check, one recovers for the non-singlet quark evolution the Curci-Furmansky-Petronzio relation~\refeq{violation}.
Other features related to the Low, Burnett, Kroll theorems~\cite{Low:1958sn} are discussed in~\cite{Dokshitzer:2005bf}.
A successfull three loop check using the $\gamma_T$ evaluated by Drell-Levy-Yan analytic continuation is 
described in~\cite{Mitov:2006ic} for the non-singlet QCD anomalous dimensions.

\subsection{Moch-Vermaseren-Moch relations and reciprocity of the kernel ${\cal P}$}

An important test of \refeq{REE} can be done in the $x\to 1$ limit. To explain it, let us briefly recall 
what are known as the Moch-Vermaseren-Moch (MVV) relations for twist-2 anomalous dimensions in QCD.
The large spin $N$ expansion of the (unpolarized) 3 loops anomalous dimensions~\cite{Moch:2004pa,Vogt:2004mw} starts with 
a leading logarithm behavior $2\,\Gamma_{\rm cusp}(\alpha_s)\,\log\,N$. The subleading $\sim \log^p\,N/N^q$ corrections
are found to obey special relations first investigated by MVV in~\cite{Moch:2004pa,Vogt:2004mw} (see also, at two loops,~\cite{Curci:1980uw}). 
Roughly speaking, these relations  predict the three loop $1/N$ terms in terms of the $N^0$ two loop ones.

Neglecting effects due to the running couplings, one immediately derives from \refeq{REE} the non-linear relation 
(after a rescaling of ${\cal P}$)
\be
\gamma_\sigma(N) = {\cal P}\left(N-\frac{1}{2}\,\sigma\,\gamma_\sigma(N)\right).
\ee
In the spirit of the derivation of the reciprocity respecting evolution equation \refeq{REE} it is natural to
guess that the kernel ${\cal P}$ obeys the Gribov-Lipatov reciprocity relation
\be
\label{eq:calprec}
{\cal P}(x) = -x\,{\cal P}(1/x).
\ee
As an immediate corollary, the following general parametrization of the large $N$ expansion of $\gamma_\sigma$ (we define $\overline N = N\,e^{\gamma_{\rm E}}$
and $A = 2\,\Gamma_{\rm cusp}$)
\be
\gamma_\sigma(N) = A\,\log\,\overline{N} + B + C_\sigma \,\frac{\log\,\overline{N}}{N} + \left(D_\sigma+\frac{1}{2}\,A\right)\frac{1}{N} + \cdots, 
\ee
must satisfy the constraints
\be
\label{eq:mvv1}
C_\sigma = -\frac{1}{2}\,\sigma\,A^2, \qquad 
D_\sigma = -\frac{1}{2}\,\sigma\,A\,B,
\ee
which are highly non-trivial since $A, B, C$ and $D$ are functions of the gauge coupling.
The first relation in (\ref{eq:mvv1}) is indeed verified at three loops by the explicit evaluation of $\gamma_\sigma$. 
The second (subleading) relation requires, in QCD, a correction related to the non-zero value of the $\beta$ function, as discussed in~\cite{Basso:2006nk}.
For twist-2 operators in the finite ${\cal N}=4$ SYM theory, it is correct as it stands.

Thus, the two MVV relations in \refeq{mvv1} strongly suggest that the reciprocity relation \refeq{calprec} holds.
In $N$-space, it is equivalent to the claim that the kernel ${\cal P}(N)$ has a large $N$ expansion in {\em integer powers of } $J^2$ of the form 
\be
\label{eq:firstreciprocity}
{\cal P}(N) = \sum_n\frac{a_n(\log\,J)}{J^{2n}},
\ee
where $J^2 = N\,(N+1)$, and $a_n$ are polynomials which can be computed in perturbation theory as series in $\alpha_s$.
The expansion \refeq{firstreciprocity} can be read as a parity invariance under $N\to -N-1$, although this must be considered only around $N=\infty$ and not
in strict sense  because of the Regge poles at negative $N$.

The property \refeq{firstreciprocity}, or its equivalent form \refeq{calprec}, has indeed been checked at three loops in~\cite{Basso:2006nk}  
for several classes of twist-2 operators in QCD.
It generates an infinite set of MVV-like relations for all the subleading terms in the large $N$ expansion of the anomalous
dimensions. The previous relations \refeq{mvv1} are just the first cases.

\section{Generalized reciprocity in ${\cal N}=4$ SYM}
\label{sec:n4}

A generalization of \refeq{firstreciprocity} has also been proposed
based on the analysis~\cite{Korchemsky:1996kh,Belitsky:2006en} of the one-loop anomalous dimensions of maximal helicity quasipartonic operators~\cite{quasipart}
built with (collinear) twist-1 fundamental fields $X$ (scalars, gauginos or gauge fields) and light-cone projected covariant derivatives. 

Such operators can be written in a general non-local form as
\be
\label{eq:O}
\mathbb{O}(z_1, \dots, z_L) = \mbox{Tr}\big\{X(z_1\,n)\cdots X(z_L\,n)\big\},
\ee
where $z\,n^\mu$ is the light-like ray and $X$ can be a (suitable) ${\cal N}=4$ scalar field $\varphi$, gaugino component $\lambda$, or 
holomorphic combination $A$ of the physical gauge field $A^\mu_\perp$. We shall denote generically such operators as $\mathbb{O}^\varphi$, 
$\mathbb{O}^\lambda$, and $\mathbb{O}^A$.
Linear combinations of such local fields provide eigenstates of the dilatation operator. 

At one-loop, these operators do not mix and transform under the collinear conformal group as $[s]^{\otimes L}$ where $[s]$ is the 
infinite dimensional $\mathfrak{sl}(2)$ representation with collinear spin 
\be
\label{eq:collspin}
s(\varphi) = \frac 1 2,\qquad
s(\psi) = 1,\qquad
s(A) = \frac 3 2.
\ee
At more than one loop, the operators $\mathbb{O}^\varphi$ and $\mathbb{O}^\lambda$ continue to scale autonomously. The reason is that $\mathbb{O}^\varphi$
belongs to the ${\cal N}=4$ $\mathfrak{sl}(2)$ subsector which is closed at all orders. Also, $\mathbb{O}^\lambda$ appears in the closed $\mathfrak{sl}(2|1)$
subsector where there is mixing between scalars and fermions, but not for the maximally fermionic component~\cite{Belitsky:2007zp}.
In the case of $\mathbb{O}^A$, the description as a {\em gluonic} operator is only correct at one-loop~\cite{Belitsky:2008wj} with mixing effects at 
higher orders (see the discussion in~\cite{Beccaria:2007pb}).

Let us now illustrate the correct extension of 
\refeq{firstreciprocity} valid in the ${\cal N}=4$ context for the operators (\ref{eq:O}).
Since the $\beta$ function is identically zero, the kernel ${\cal P}(N)$ for the space-like ordinary anomalous dimensions 
obeys the relation
\be
\label{eq:P}
\gamma(N) = {\cal P}\left(N+\frac 1 2 \gamma(N)\right).
\ee
The one-loop anomalous dimensions of $\mathbb{O}^{\varphi,\lambda, A}$ can be computed as energies of $\mbox{XXX}_{-s}$ integrable chains
and in particular can be studied at large Lorentz spin. The analysis of~\cite{Korchemsky:1996kh,Belitsky:2006en} suggests 
that reciprocity takes the form 
\be
\label{eq:reciprocity}
{\cal P}(N) = \sum_n\frac{a_n(\log\,J)}{J^{2n}},
\ee
where $J$ is obtained by replacing $N(N+1)$ with the suitable Casimir of the 
collinear conformal subgroup $SL(2, \mathbb{R})\subset SO(4,2)$
\be
\label{eq:J}
J^2 = (N+L\,s-1)\,(N+L\,s).
\ee
If the expansion (\ref{eq:reciprocity}) holds, we shall say that ${\cal P}$
is a {\em reciprocity respecting} (RR) kernel.
Beyond one loop, a test of reciprocity requires the knowledge of the multi-loop anomalous dimensions as closed functions of $N$. These are currently available in the cases of twist-2 and 3, 
as discussed in the following sections.

\subsection{Twist-2 universal supermultiplet}

As a first example, we discuss twist-2 operators. Due to supersymmetry, the collinear conformal spin (\ref{eq:collspin}) is irrelevant and we can consider 
the simplest case of operators built with scalar fields. These are described by non-nested $\mathfrak{sl}(2)$ Bethe equations~\cite{Staudacher:2004tk}. In this case,
we have, as in QCD,  
\be
J^2 = N\,(N+1).
\ee
Let us briefly recall how the reciprocity property \refeq{reciprocity} of a generic function $f(N)$ translates into 
the GL reciprocity of its Mellin transform $F(z)$ defined by 
\be
f(N) = \int_0^1\frac{dz}{z}\,z^N\,F(z).
\ee
This is a useful exercise since we shall generalize it to other cases later.
Here, we follow the method by~\cite{Basso:2006nk}. With the change of variable
\be
z=e^{-\lambda\,x},\qquad \lambda = \left(J^2+\frac{1}{4}\right)^{-1/2} = \frac{1}{N+\frac{1}{2}},
\ee
we can write
\be
f(N) = \lambda\,\int_0^\infty dx\,e^{-x}\,e^{\lambda\,x/2}\,F(e^{-\lambda\,x}).
\ee
Reciprocity means that 
the integrand is locally odd under $\lambda\to -\lambda$ in a neighborhood of $\lambda=0$.
This gives 
\be
e^{\lambda\,x/2} \,F(e^{-\lambda\,x}) = -e^{-\lambda\,x/2} \, F(e^{\lambda\,x}),
\ee
which means
\be
F(z) = -z\,F(1/z).
\ee
In the paper~\cite{Dokshitzer:2006nm}, this relation is proved for the known three loops anomalous dimensions derived 
by the Kotikov, Lipatov, Onishchenko and Velizhanin (KLOV)  maximum transcendentality principle~\cite{Kotikov:2004er}.
The method exploits several properties of the nested harmonic sums (see App.~(\ref{app:harmonic})) which are the building block for the 
perturbative result.
The same conclusion is also obtained in~\cite{Basso:2006nk} by directly checking the expansion \refeq{reciprocity}.

\subsection{Twist-3 operators with scalar fields}

Again, these are described by non-nested $\mathfrak{sl}(2)$ Bethe equations. We have 
\be
J^2 = 4\,\frac{N}{2}\,\left(\frac{N}{2}+1\right)+\frac{3}{4}.
\ee
The constant $3/4$ is irrelevant to the proof of reciprocity and one can define
\be
J^2 \stackrel{def}{=} \frac{N}{2}\,\left(\frac{N}{2}+1\right).
\ee
Four loops closed expressions for $\gamma(N)$ have been obtained in~\cite{Beccaria:2007cn,Kotikov:2007cy}.
They involve harmonic sums evaluated at
\be
\widetilde{N} = \frac{N}{2}.
\ee
Since $J^2 = \widetilde{N}(\widetilde{N}+1)$, the reciprocity proof can be done with the methods used in the twist-2
case with scalar fields. This calculation is done in~\cite{Beccaria:2007bb}.

\subsection{Twist-3 operators with gauginos}

This case has been treated in~\cite{Beccaria:2007vh} where the following result was obtained
\be
\gamma_{\lambda\lambda\lambda}(N) = \gamma_{\varphi\varphi}(N+2).
\ee
Here, $\gamma_{\lambda\lambda\lambda}$ is the anomalous dimension in this sector and $\gamma_{\varphi\varphi}$ is the one 
for twist-2 operators with scalar fields.

>From the two relations
\ba
\gamma_{\varphi\varphi}(N) &=& {\cal P}_{\varphi\varphi}\left(N+\frac{1}{2}\gamma_{\varphi\varphi}(N)\right), \\
\gamma_{\lambda\lambda\lambda}(N) &=& {\cal P}_{\lambda\lambda\lambda}\left(N+\frac{1}{2}\gamma_{\lambda\lambda\lambda}(N)\right), 
\ea
we deduce
\be
{\cal P}_{\lambda\lambda\lambda}(N) = {\cal P}_{\varphi\varphi}(N+2).
\ee
Since ${\cal P}_{\varphi\varphi}(N)$ is reciprocal with respect to $J^2 = N(N+1)$ we conclude that ${\cal P}_{\lambda\lambda\lambda}(N)$ is 
reciprocal with respect to 
\be
J^2 = (N+2)(N+3).
\ee
This is precisely the Casimir in this sector ($L=3$, $s=1$), showing that again reciprocity is respected.

\subsection{Twist-3 operators with gauge fields}

For this case, three loop anomalous dimensions are known and a few MVV relations have been tested~\cite{Beccaria:2007pb}. 
Since we are going to extend the calculation and the reciprocity proof to the more difficult four loop case, we devote to this sector the next Section.

\section{Gluonic operators}
\label{sec:gluonic}

As we mentioned in Section 3, we are interested in single-trace maximal helicity quasi-partonic operators which in the light-cone gauge take
the form 
\be
\label{eq:gaugeops}
\mathbb{O}_{N, L}^A = \sum_{n_1+\cdots n_L=N} a_{n_1,\dots n_L}\,\mbox{Tr}\left\{\partial_+^{n_1} A(0)\cdots \partial_+^{n_L} 
A(0)\right\},\ \ n_i\in\mathbb{N},
\ee
where $A$ is the holomorphic combination of the physical gauge degrees of freedom $A^\mu_\perp$ and $\partial_+$ is the light-cone 
projected covariant derivative (in light-cone gauge the gauge links are absent).
The coefficients $\{a_\mathbf{n}\}$ are such that $\mathbb{O}^A_{N, L}$ is a scaling field, eigenvector of the dilatation
operator. The total Lorentz spin is $N=n_1 + \cdots n_L$. The number of elementary fields equals the twist 
$L$, {\em i.e.} the classical dimension minus the Lorentz spin.

At one-loop, the anomalous dimensions of the above operators can be found from the spectrum of a 
non-compact $XXX_{-3/2}$ spin chain with $L$ sites. 
At higher orders we abandon the quasipartonic detailed description and work in terms of superconformal multiplets. 
The first step is to identify the $\mathfrak{psu}(2,2|4)$ primary of the multiplet where such operators appear as descendant. 
In full generality such multiplets in twist-3
appear in the decomposition of the symmetric triple tensor product $(V_F\otimes V_F\otimes V_F)_S$
where $V_F$ is the singleton infinite dimensional irreducible representation of  $\mathfrak{psu}(2,2|4)$.

Following~\cite{Beisert:2004di}, we have a detailed decomposition 
\be
(V_F\otimes V_F\otimes V_F)_S = \mathop{\bigoplus_{n=0}}_{k\in\mathbb{Z}}^\infty c_n\left[V_{2k, n}+V_{2k+1, n+3}\right],
\ee
where $c_n$ are suitable multiplicities and $V_{n,m}$ well defined modules. In particular, 
for even $N$ and $m=2$, the one-loop lowest anomalous dimension in $V_{2,N}$ is associated with an unpaired state and has 
been proposed to be~\cite{Beisert:2004di} 
\be
\label{eq:oneloopmodule}
\gamma_{2, N} =  \frac{\lambda}{8\pi^2}\left[
2\,S_1\left(\frac{N}{2}+1\right)
+2\,S_1\left(\frac{N}{2}+2\right)+4\right]= \frac{\lambda}{8\pi^2}\left[
2\,S_1\left(\frac{N}{2}+1\right)+\frac{4}{N+4}+4\right],
\ee
where $g^2 = \lambda/(8\,\pi^2) = g_{\rm YM}^2\,N_c/(8\,\pi^2)$ is the scaled 't Hooft coupling, 
fixed in the planar $N_c\to\infty$ limit.
This result is in agreement with the analysis of maximal helicity 3 gluon operators in QCD~\cite{Belitsky:1999bf} and identifies
the module  $V_{2,N}$ with the one containing the considered operators.
The second expression in (\ref{eq:oneloopmodule}) 
fully reveals the {\em violation of the maximum transcendentality principle}~\cite{Kotikov:2004er}, a novel feature of the gauge sector already discussed in~\cite{Beccaria:2007pb}.

\subsection{Long-range Bethe equations}

The long-range (asymptotic) Bethe equations for the full $\mathfrak{psu}(2,2|4)$ theory have been 
formulated in~\cite{Beisert:2005fw} in 4 equivalent forms. The most convenient one 
has the following degree assignment for the module $V_{2,N}$
\be
\begin{minipage}{260pt}
\setlength{\unitlength}{1pt}
\small\thicklines
\begin{picture}(260,55)(-10,-30)
%
\put(-32,0){\line(1,0){22}}  
\put(  0,00){\circle{15}}
\put( -5,-5){\line(1, 1){10}}  
\put( -5, 5){\line(1,-1){10}}  
\dottedline{3}(8,0)(32,0)    
\put( 40,00){\circle{15}}     
\dottedline{3}(48,0)(72,0)   
\put( 80,00){\circle{15}}
\put( 75,-5){\line(1, 1){10}}  
\put( 75, 5){\line(1,-1){10}}  
\put( 80,-15){\makebox(0,0)[t]{$N+3$}}  
\put( 87,00){\line(1,0){26}} 
\put(120,00){\circle{15}}
\put(120,15){\makebox(0,0)[b]{$+1$}} 
\put(120,-15){\makebox(0,0)[t]{$N+4$}} 
\put(127,00){\line(1,0){26}} 
\put(160,00){\circle{15}}
\put(155,-5){\line(1, 1){10}}  
\put(155, 5){\line(1,-1){10}}  
\put(160,-15){\makebox(0,0)[t]{$N+2$}} 
\dottedline{3}(168,0)(192,0) 
\put(200,00){\circle{15}}
\put(200,-15){\makebox(0,0)[t]{$1$}} 
\dottedline{3}(208,0)(232,0) 
\put(240,00){\circle{15}}
\put(235,-5){\line(1, 1){10}} 
\put(235, 5){\line(1,-1){10}} 
\put(250,0){\line(1,0){20}} 
\end{picture}
\end{minipage}
\ee
A detailed description of the perturbative solution of the associated Bethe equation has already been illustrated in~\cite{Beccaria:2007pb}.
The only new ingredient at four loops is the dressing phase which we have taken from~\cite{Beisert:2006ez}. It gives a 
contribution which in the notation of that paper can be written
\be
\gamma_4 = \gamma_4^{\rm no\ dressing} + \beta\,\gamma_4^{\rm dressing}.
\ee
The correct value is $\beta = \zeta_3$. As we shall discuss, the dressing contribution is separately reciprocity respecting, precisely
as it happens in the case of twist-3 operators built with scalar fields~\cite{Beccaria:2007bb}. Therefore, we shall keep it separate in the following discussion.

\subsection{Three loop results}

The results obtained in~\cite{Beccaria:2007pb} at three loops can be summarized in the following closed expressions
\ba
\label{eq:gamma1}
\gamma_1 &=& 4\,S_1+\frac{2}{n+1}+4, \\
\label{eq:gamma2}
\gamma_2 &=& -2\,S_3-4\,S_1\,S_2-\frac{2\,S_2}{n+1}-\frac{2\,S_1}{(n+1)^2}-\frac{2}{(n+1)^3} + \\
&& -4\,S_2-\frac{2}{(n+1)^2}-8, \nonumber \\
\label{eq:gamma3}
\gamma_3 &=& 
+5\,S_5
+6\, S_2\, S_3
-4\, S_{2,3}
+4\, S_{4,1}
-8\,S_{3,1,1} \\
&&
+\left(4\, S_2^2+2\,S_4+8\,S_{3,1}\right)\,S_1 \nonumber \\
&&
+\frac{-S_4+4\,S_{2,2}+4\, S_{3,1}}{n+1}
+\frac{4\, S_1\, S_2+S_3}{(n+1)^2}
+\frac{2\, S_1^2+3\, S_2}{(n+1)^3} \nonumber \\
&&
+\frac{6\, S_1}{(n+1)^4}
+\frac{4}{(n+1)^5} \nonumber \\
&&
-2 \,S_4
+8\,S_{2,2} 
+8\, S_{3,1} \nonumber \\
&&
+\frac{4\, S_2}{(n+1)^2}
+\frac{4\, S_1}{(n+1)^3}
+\frac{6}{(n+1)^4} \nonumber \\ 
&&
+ 8\, S_2
+32, \nonumber
\ea
where $n = \frac{N}{2}+1$ and $S_a\equiv S_a(n)$ are nested harmonic sums (see App.~(\ref{app:harmonic})).

\subsection{Some structural properties}
\label{sec:structural}

Before attacking the problem of deriving a four loop expression for the anomalous dimension, it is convenient to pause 
and illustrate some structural properties of the three loop result. 
The general form of $\gamma_n$ obeys at three loops the generalized KLOV structure
\ba
\label{eq:Ansatz}
\gamma_n &=& \sum_{\tau=0}^{2\,n-1}\gamma_n^{(\tau)}, \\
\gamma_n^{(\tau)} &=& \sum_{k+\ell = \tau}\frac{{\cal H}_{\tau,\ell}(n)}{(n+1)^k}, \nonumber
\ea
where ${\cal H}_{\tau,\ell}(n)$ is a combination of harmonic sums with homogeneous fixed transcendentality $\ell$. The terms with $k=0$ have maximum \tran, all the others have subleading \tran. 
Some structural properties that emerge are the following.

\begin{enumerate}
\item {\bf $\mathfrak{sl}(2)$ limit}. The maximum \tran terms without $1/(n+1)$ factors are those 
already computed in the sector with $L=3$ and scalar fields~\cite{Beccaria:2007cn,Kotikov:2007cy}
\be
\label{eq:sl2limit}
{\cal H}_{2\,n-1, 2\,n-1} = \mbox{identical to $L=3$, $s=1/2$ sector}.
\ee

\item {\bf Minimal transcendentality 1 terms}. With the exception of $\gamma_1$ we have 
\be
\label{eq:cuspmatch}
\gamma_n^{(1)} = 0.
\ee

\item {\bf Inheritance}. Write the maximum \tran ${\cal H}_{2\,n-1, 2\,n-1}(n)$ in the canonical basis of harmonic functions (see App.~(\ref{app:harmonic})).
Consider the expression
\be
\frac{1}{2}\left[{\cal H}_{2\,n-1, 2\,n-1}(n) + {\cal H}_{2\,n-1, 2\,n-1}(n+1)\right],
\ee
and expand the second term using recursively the relations
\be
\label{eq:rule}
S_{a, \mathbf{b}}(n+1) \longrightarrow \frac{\rho}{(n+1)^a}\,S_{\mathbf{b}}(n+1)+S_{a, \mathbf{b}}(n),
\ee
where $\rho$ is an auxiliary counting variable. When the process of expansion is completed we have 
\be
\frac{1}{2}\left[{\cal H}_{2\,n-1, 2\,n-1}(n) + {\cal H}_{2\,n-1, 2\,n-1}(n+1)\right] =  {\cal H}_{2\,n-1, 2\,n-1}(n) +
\ee
$$ 
+  \sum_{|\mathbf{a}|+k = 2\,n-1} P_{\mathbf{a}, k}(\rho) \frac{S_{\mathbf{a}}(n)}{(n+1)^k},
$$
where
\be
\mathbf{a} = \{a_1, \dots, a_p\}\ \longrightarrow\ |\mathbf{a}| = a_1 + \cdots + a_p,
\ee
and $P_{\mathbf{a}, k}(\rho)$ is a polynomial. Then, we have 
\be
\gamma_n^{(2\,n-1)} =  {\cal H}_{2\,n-1, 2\,n-1}(n) + \mathop{\sum_{|\mathbf{a}|+k = 2\,n-1}}_{P_{\mathbf{a}, k}(\rho)\ \rm linear} \frac{S_{\mathbf{a}}(n)}{(n+1)^k} + 
\ee
$$
+\mathop{\sum_{|\mathbf{a}|+k = 2\,n-1}}_{P_{\mathbf{a}, k}(\rho)\ \rm non linear} c_{\mathbf{a}, k} \frac{S_{\mathbf{a}}(n)}{(n+1)^k},
$$
where $c_{\mathbf{a}, k}$ are undetermined constants. This inheritance principle fixes many of the maximum \tran terms of $\gamma_n$. The terms with undetermined
coefficients are in any case a subset of all the possible terms.

\end{enumerate}

Let us illustrate two examples of the inheritance property. At one-loop, we start from the $\mathfrak{sl}(2)$ result
\be
\gamma_1^{\mathfrak{sl}(2)} = 4\,S_1,
\ee
and consider the sum
\be
\frac{1}{2}\left[4\,S_1(n)+4\,S_1(n+1)\right].
\ee
Expanding using the rule \refeq{rule}, we find
\be
4\,S_1(n)+\frac{2\,\rho}{n+1}.
\ee
Thus, inheritance fully predicts the \tran 1 terms 
\be
4\,S_1(n)+\frac{2}{n+1},
\ee
in agreement with \refeq{gamma1}.

At two loops, we start from the $\mathfrak{sl}(2)$ result that we write in canonical form 
\be
\gamma_2^{\mathfrak{sl}(2)} = -2\,S_3-4\,S_1\,S_3 = -4\,S_{1,2}-4\,S_{2,1}+2\,S_3,
\ee
and consider the sum
\be
\frac{1}{2}\left[-4\,S_{1,2}(n)-4\,S_{2,1}(n)+2\,S_3(n)-4\,S_{1,2}(n+1)-4\,S_{2,1}(n+1)+2\,S_3(n+1)\right].
\ee
This gives back the $\mathfrak{sl}(2)$ result computed at $n$ plus various induced terms that are
\be
-\frac{2\,\rho}{n+1}\,S_2(n)-\frac{2\,\rho}{(n+1)^2}\,S_1(n)+\rho(1-4\,\rho)\frac{1}{(n+1)^3}.
\ee
The prediction from inheritance is now
\be
-\frac{2}{n+1}\,S_2(n)-\frac{2}{(n+1)^2}\,S_1(n)+\frac{c}{(n+1)^3},
\ee
where $c$ is an undetermined constant. Without resorting to the inheritance property, we should have needed four coefficients for the possible allowed terms
\be
\frac{S_2}{n+1},\quad \frac{S_{1,1}}{n+1},\quad \frac{S_1}{(n+1)^2},\quad \frac{1}{(n+1)^3}. 
\ee

\section{The four loop anomalous dimension}
\label{sec:fourloops}

We have computed a long list of values of $\gamma_4(n)$ as exact rational numbers obtained from the perturbative 
expansion of the long-range Bethe equations. We have matched them against the general Ansatz \refeq{Ansatz}. A very large number of possible terms appear with unknown
coefficients. To reduce them, we have imposed the {\em inheritance} property described in Sec.~(\ref{sec:structural}) as well as the condition \refeq{cuspmatch}.
The resulting reduced Ansatz matches the list $\{\gamma_4(n)\}$ with rather simple integer coefficient. Our list is longer than the number of coefficients and 
we checked that it is perfectly reproduced. Also, we extended the list to even larger values of $n$ where we only have a (very long) decimal approximation to $\gamma_4(n)$
again in agreement with the solution found.

We use the notation of \refeq{Ansatz} to present our result. 
We begin with the non-dressing contributions to $\gamma_4$. The terms with maximal \tran are 

\ba
{\cal H}_{7,7} &=& 
\frac{S_7}{2}+7 \,S_{1,6}+15 \,S_{2,5}-5 \,S_{3,4}-29 \,S_{4,3}-21 \,S_{5,2}-5
   \,S_{6,1}-40 \,S_{1,1,5}-32 \,S_{1,2,4}+24 \,S_{1,3,3}+  \nonumber \\
&& + 32 \,S_{1,4,2}-32 \,S_{2,1,4}+20
   \,S_{2,2,3}+40 \,S_{2,3,2}+4 \,S_{2,4,1}+24 \,S_{3,1,3}+44 \,S_{3,2,2}+24
   \,S_{3,3,1}+ \nonumber \\
&& + 36 \,S_{4,1,2}+36 \,S_{4,2,1}+24 \,S_{5,1,1}+80 \,S_{1,1,1,4}-16
   \,S_{1,1,3,2}+32 \,S_{1,1,4,1}-24 \,S_{1,2,2,2}+16 \,S_{1,2,3,1} +  \nonumber \\
&& -24 \,S_{1,3,1,2}-24
   \,S_{1,3,2,1}-24 \,S_{1,4,1,1}-24 \,S_{2,1,2,2}+16 \,S_{2,1,3,1}-24 \,S_{2,2,1,2}-24
   \,S_{2,2,2,1} +  \nonumber \\
&& -24 \,S_{2,3,1,1}-24 \,S_{3,1,1,2}-24 \,S_{3,1,2,1}-24 \,S_{3,2,1,1}-24
   \,S_{4,1,1,1}-64 \,S_{1,1,1,3,1},  \nonumber \\
{\cal H}_{7,6} &=&
\frac{7 S_6}{2}-20 \,S_{1,5}-16 \,S_{2,4}+12 \,S_{3,3}+16 \,S_{4,2}+40 \,S_{1,1,4}-8
   \,S_{1,3,2}+16 \,S_{1,4,1}-12 \,S_{2,2,2}+8 \,S_{2,3,1} + \nonumber \\
&& -12 \,S_{3,1,2}-12
   \,S_{3,2,1}-12 \,S_{4,1,1}-32 \,S_{1,1,3,1}, \nonumber \\
{\cal H}_{7,5} &=&
-\frac{15 \,S_5}{2}+14 \,S_{1,4}+10 \,S_{2,3}+14 \,S_{3,2}+14 \,S_{4,1}-12 \,S_{1,2,2}-16
   \,S_{1,3,1}-12 \,S_{2,1,2}+ \nonumber\\
&& -12 \,S_{2,2,1}-12 \,S_{3,1,1}, \nonumber \\
{\cal H}_{7,4} &=&
-\frac{3 \,S_4}{2}+12 \,S_{1,3}+4 \,S_{2,2}+4 \,S_{3,1}-12 \,S_{1,1,2}-12 \,S_{1,2,1}-12
   \,S_{2,1,1}, \nonumber \\
{\cal H}_{7,3} &=&
11 \,S_3-9 \,S_{1,2}-9 \,S_{2,1}-12 \,S_{1,1,1}, \nonumber \\
{\cal H}_{7,2} &=&
4 \,S_2-24 \,S_{1,1}, \nonumber \\
{\cal H}_{7,1} &=&
-\frac{39 \,S_1}{2}, \nonumber \\
{\cal H}_{7,0} &=&
-\frac{39}{4}. 
\ea
The other terms with lower \tran read
\ba
{\cal H}_{6,6} &=&
7 \,S_6-40 \,S_{1,5}-32 \,S_{2,4}+24 \,S_{3,3}+32 \,S_{4,2}+80 \,S_{1,1,4}-16 \,S_{1,3,2}+32
   \,S_{1,4,1}-24 \,S_{2,2,2}+ \nonumber \\
&& +16 \,S_{2,3,1}-24 \,S_{3,1,2}-24 \,S_{3,2,1}-24
   \,S_{4,1,1}-64 \,S_{1,1,3,1}, \nonumber \\
{\cal H}_{6,5} &=&
-20 \,S_5+40 \,S_{1,4}-8 \,S_{3,2}+16 \,S_{4,1}-32 \,S_{1,3,1}, \nonumber \\
{\cal H}_{6,4} &=&
10 \,S_4+4 \,S_{1,3}-12 \,S_{2,2}-12 \,S_{3,1}, \nonumber \\
{\cal H}_{6,3} &=&
14 \,S_3-12 \,S_{1,2}-12 \,S_{2,1}, \nonumber \\
{\cal H}_{6,2} &=&
-9 \,S_2-12 \,S_{1,1}, \nonumber \\
{\cal H}_{6,1} &=&
-22 \,S_1, \nonumber \\
{\cal H}_{6,0} &=&
-\frac{37}{2}, \nonumber \\
{\cal H}_{5,5} &=&
-20 \,S_5+40 \,S_{1,4}-8 \,S_{3,2}+16 \,S_{4,1}-32 \,S_{1,3,1}, \nonumber \\
{\cal H}_{5,4} &=&
20 \,S_4-16 \,S_{3,1}, \nonumber \\
{\cal H}_{5,3} &=&
4 \,S_3, \nonumber \\
{\cal H}_{5,2} &=&
-4 \,S_2, \nonumber \\
{\cal H}_{5,1} &=&
0, \nonumber \\
{\cal H}_{5,0} &=&
-2, \nonumber \\
{\cal H}_{4,4} &=&
20 \,S_4-16 \,S_{2,2}-32 \,S_{3,1}, \nonumber \\
{\cal H}_{4,3} &=&
0, \nonumber \\
{\cal H}_{4,2} &=&
-4 \,S_2, \nonumber \\
{\cal H}_{4,1} &=&
-8 \,S_1, \nonumber \\
{\cal H}_{4,0} &=&
2, \nonumber \\
{\cal H}_{3,1} &=&
-8 \,S_1, \nonumber \\
{\cal H}_{3,0} &=&
4, \nonumber \\
{\cal H}_{2,2} &=&
-32 \,S_2, \nonumber  \\
{\cal H}_{2,1} &=&
0 ,\nonumber \\
{\cal H}_{2,0} &=&
8 ,\nonumber \\
{\cal H}_{0,0} &=& 
-160.
\ea
Finally, the dressing contribution reads
\ba
\label{eq:dressing4}
\gamma_4^{\rm dressing} &=&
-8 \,S_3 \,S_1-\frac{8 S_1}{(n+1)^2}-\frac{4 S_1}{(n+1)^3}-\frac{4 S_3}{n+1}-8
   S_3 + \nonumber \\
&& -\frac{8}{(n+1)^2}-\frac{8}{(n+1)^3}-\frac{2}{(n+1)^4} ,
\ea
and, multiplied by $\zeta_3$, consistently shows the proper generalized \tran. 

It is not difficult to immediately check that the correct  universal cusp anomalous dimension $\Gamma_{\rm cusp}(g)$ at four 
loops is reproduced by the leading large $N$ expansion of the formulas above.  While all the terms with $1/(n+1)$ factors are 
suppressed in the large $N$ limit, those with maximum \tran and without those factors are in fact the same as in the $L=3$ scalar 
sector, where it has been already checked~\cite{Beccaria:2007cn} that
\be
\label{eq:cusp}
\gamma_4^{\rm no\ dressing} + \zeta_3\,\gamma_4^{\rm dressing} = -\left(\frac{73\,\pi^6}{630}+4\zeta_3^2\right)\,\log N + {\cal O}(N^0),\qquad {\rm at} ~~N\to\infty.
\ee

\section{Proof of reciprocity}
\label{sec:proof}

This section contains the complete proof of reciprocity of the four loop anomalous dimension. It is organized as follows.
In Sec.~(\ref{subsec:mellin}) we derive the correct reciprocity in Mellin $x$-space for this sector.
In Sec.~(\ref{sec:construction}) we present some useful technical result.
In Sec.~(\ref{subsec:reduction}) we illustrate a reduction algorithm to write explicitly and in an automatic way 
the separately reciprocity respecting structures. Finally, in
Sec.~(\ref{subsec:final}) we collect the results.

\subsection{Reciprocity condition from Mellin transformation}
\label{subsec:mellin}

The quadratic Casimir is 
\be
J^2 = N^2 + 8\,N + \frac{63}{4} = 4\,n(n+2)+\frac{15}{4}.
\ee
The {\em effective} $J^2$ can be defined as
\be
J^2 \stackrel{def}{=} n\,(n+2).
\ee
Let us consider now the Mellin transformation of a function which is expressed as depending on $n$
\be
f(n) = \int_0^1\frac{dz}{z}\,z^n\,F(z),
\ee
Let us define
\be
z=e^{-\lambda\,x},\qquad \lambda = \left(J^2+1\right)^{-1/2} = \frac{1}{n+1},
\ee
and write
\be
f(n) = \lambda\,\int_0^\infty dx\,e^{-x}\,e^{\lambda\,x}\,F(e^{-\lambda\,x}).
\ee
The absence of half-integer powers of $J^2$ at large $n$ is equivalent to the requirement that the integrand is 
locally odd under $\lambda\to -\lambda$ in a neighborhood of $\lambda=0$.
This gives 
\be
e^{\lambda\,x} \,F(e^{-\lambda\,x}) = -e^{-\lambda\,x} \, F(e^{\lambda\,x}),
\ee
or
\be
F(z) = -z^2\,F(z^{-1}).
\ee
>From this, a useful theorem follows
\begin{theorem}
\label{thm:rec}
Let $f(n)$ be reciprocal with respect to $J^2 = n(n+1)$. Then, the combination 
\be
\widetilde{f}(n) = f(n) + f(n+1),
\ee
is reciprocal with respect to $J^2 = n(n+2)$.
\end{theorem}

{\em Proof:} We simply write
\ba
\widetilde{f}(n) &=& f(n) + f(n+1) = \int_0^1\frac{dz}{z}\,z^n\,F(z) + \int_0^1\frac{dz}{z}\,z^{n+1}\,F(z) = \\
&=& \int_0^1\frac{dz}{z}\,z^n\,(z+1)\,F(z),
\ea
which means
\be
\widetilde{F}(z) = (z+1)\,F(z).
\ee
Using now $F(z) = -z\,F(1/z)$ we find
\be
\widetilde{F}(z) = (z+1)\,F(z) = -z\,(z+1)\,F(1/z) = -z^2\,\widetilde{F}(1/z).
\ee
\hfill $\Box$

This theorem can be used as follows. We compute the four loop function ${\cal P}(N)$ and express it in terms of $n=\frac{N}{2}+1$.
Then we rewrite it using symmetric combinations of harmonic terms which are reciprocal with respect to $n(n+1)$. These have been
classified and listed in~\cite{Beccaria:2007bb}. The next section summarizes what we need.

\subsection{Reciprocity respecting combinations with respect to $n(n+1)$}
\label{sec:construction}

Let us consider the following {\bf linear map} defined on linear combinations of simple $S$ sums by 
\be
\Phi_a (S_{b, \mathbf{c}}) = S_{a, b, \mathbf{c}}-\frac{1}{2}\, S_{a+b, \mathbf{c}}.
\ee
Define also 
\ba
I_a &=& S_a, \\
I_{a_1, a_2, \dots, a_n} &=& \Phi_{a_1}(\Phi_{a_2}(\cdots\, \Phi_{a_{n-1}}(S_{a_n})\cdots ).
\ea
For instance,
\be
I_{a,b} = S_{a,b}-\frac{1}{2} S_{a+b}.
\ee
Then, we have the following important result
\begin{theorem}[\cite{Beccaria:2007bb}]
\label{thm:odd}
The combinations $I_{a_1, \dots, a_n}$ with odd $a_1, \dots, a_n$ have a large $N$ reciprocity respecting expansion
\be
I_{a_1, \dots, a_n} = \sum_{\ell=0}^\infty \frac{P_\ell(\log\,J^2)}{J^{2\ell}},
\ee
where $J^2 = N(N+1)$ and $P_\ell$ is a polynomial.
\end{theorem}

\subsection{Reduction algorithm}
\label{subsec:reduction}
The general strategy to prove reciprocity is as follows. Let us consider a nested harmonic sum
$S_\mathbf{a}(n)$ with $\mathbf{a}=(a_1, \dots, a_k)$ and all $a_i$ odd. The sum $S_\mathbf{a}(n)$ is the unique
maximal depth term appearing in the expansion of the invariant $I_\mathbf{a}$ defined in Sec.~(\ref{sec:construction}).
Examples are:
\ba
I_{1,3} &=& S_{1,3}-\frac{1}{2}\,S_4, \\
I_{1,1,3} &=& S_{1,1,3}-\frac{1}{2}\,S_{2,3}-\frac{1}{2}\,S_{1,4}+\frac{1}{4}\,S_5.\nonumber
\ea
This means that we can write
\be
S_\mathbf{a}(n) = I_\mathbf{a}(n)+R_\mathbf{a}(n),\qquad \mbox{depth}(R_\mathbf{a}) < k.
\ee
>From Theorem~(\ref{thm:odd}), we know that $I_\mathbf{a}(n)$ is reciprocity respecting with respect to the combination $n\,(n+1)$. 
We then write
\be
\label{eq:decomp}
S_\mathbf{a}(n) = \frac{I_\mathbf{a}(n)+I_\mathbf{a}(n+1)}{2}+\frac{I_\mathbf{a}(n)-I_\mathbf{a}(n+1)}{2}
+R_\mathbf{a}(n).
\ee
We rename the first term 
\be
\widetilde{I}_\mathbf{a}(n) = I_\mathbf{a}(n) + I_\mathbf{a}(n+1),
\ee
and we know from Theorem~(\ref{thm:rec}) that it is reciprocity respecting with respect to the combination $n\,(n+2)$. 
Both the remaining two terms in \refeq{decomp} have depth strictly smaller than $k$. For example
\be
S_{1,3}(n) = \frac{1}{2}\,\widetilde{I}_{1,3}-\frac{1}{2}\,\frac{S_3(n)}{n+1}+\frac{1}{2}\,S_4(n)-\frac{1}{4}\frac{1}{(n+1)^4}.
\ee
The algorithm can now be iteratively applied to the generated terms of depth $k-1$. 

This strategy can be used to prove reciprocity (with respect to $n\,(n+2)$) of a generic linear combination of products of nested harmonic sums
with possible $(n+1)^{-p}$ factors. To this aim, we first combine all products of nested harmonic sums using the general shuffle algebra relation \refeq{shuffle}.
Then the algorithm is applied up to depth 0. The final result is a combination of invariants $\widetilde{I}_\mathbf{a}$ and factors $(n+1)^{-p}$. If all the indices in the 
invariants $\widetilde{I}_\mathbf{a}$ are odd and all the exponents $p$ are even, the initial expression is automatically reciprocity respecting with respect to $n\,(n+2)$.
The constraint on $p$ is due to the relation
\be
n+1 = \sqrt{n\,(n+2)+1}.
\ee

\subsection{Results for $\cal P$ at four loops}
\label{subsec:final}

The $\cal P$ function reads at four loops and in terms of $n=\frac{N}{2}+1$ ($\partial\equiv\partial_n$)
\ba
{\cal P}(n) &=& \sum_{k=1}^\infty \frac{1}{k!}\left(-\frac{1}{4}\partial\right)^{k-1}[\gamma(n)]^k = \\
&=& \gamma-\frac{1}{8}\,(\gamma^2)'+\frac{1}{96}\,(\gamma^3)''-\frac{1}{1536}\,(\gamma^4)''' + \cdots .\nonumber
\ea
Replacing the perturbative expansions
\be
{\cal P} = \sum_{k=1}^\infty {\cal P}_k\,g^{2\,k},\qquad
\gamma = \sum_{k=1}^\infty \gamma_k\,g^{2\,k},
\ee
we find
\ba
{\cal P}_1 &=& \gamma_1, \\
{\cal P}_2 &=& \gamma_2-\frac{1}{8}\,(\gamma_1^2)',\\
{\cal P}_3 &=& \gamma_3-\frac{1}{4}\,(\gamma_1\,\gamma_2)'+\frac{1}{96}\,(\gamma_1^3)'',\\
{\cal P}_4 &=& \gamma_4-\frac{1}{8}\,(\gamma_2^2+2\,\gamma_1\,\gamma_3)'+\frac{1}{32}\,(\gamma_1^2\,\gamma_2)''-\frac{1}{1536}(\gamma_1^4)'''.
\ea
These expressions can be computed taking derivatives using the results of Sec.~(\ref{sec:der}). Applying the algorithm for the reduction 
to invariants we find immediately the one-loop result in manifestly reciprocity respecting form
\be
{\cal P}_1 = 2\, \widetilde{I}_1 + 
4 .
\ee
At two loops, the same calculation gives
\be
{\cal P}_2 = 
-\,\widetilde{I}_3 
-\frac{1}{3} \pi ^2 \,\widetilde{I}_1 
-8-\frac{2 \pi ^2}{3} .
\ee
At three loops, we obtain the result
\ba
{\cal P}_3 &=&
\frac{\,\widetilde{I}_3}{2 (n+1)^2}+\frac{3 \,\widetilde{I}_5}{2}-4 \,\widetilde{I}_{1,1,3}
+\frac{2}{(n+1)^4}-4 \,\widetilde{I}_{1,3}
+\frac{\pi ^2 \,\widetilde{I}_3}{6} + \nonumber\\
&& -2 \,\widetilde{I}_3
+4 \,\widetilde{I}_{1,1} \zeta _3-\frac{\zeta _3}{(n+1)^2}-\frac{4}{(n+1)^2}
+4 \zeta _3 \,\widetilde{I}_1+\frac{4 \pi ^4 \,\widetilde{I}_1}{45} + \nonumber\\
&& +4 \zeta _3+\frac{8 \pi ^4}{45}+\frac{4 \pi ^2}{3}+32 .
\ea
Factors $1/(n+1)$ with even exponent appear and do not spoil reciprocity as discussed above.

The non-dressing four loop result is rather long but can be obtained in a straightforward way in the reciprocity respecting form
\ba
{\cal P}_4^{\rm no \ dressing} &=&
-\frac{3 \,\widetilde{I}_1}{4
   (n+1)^6}-\frac{\,\widetilde{I}_3}{(n+1)^4}-\frac{\,\widetilde{I}_5}{(n+1)^2}-\frac{13
   \,\widetilde{I}_7}{4}+8 \,\widetilde{I}_{1,1,5}+\frac{4
   \,\widetilde{I}_{1,3,1}}{(n+1)^2}+4 \,\widetilde{I}_{1,3,3}+\nonumber\\
&& + 4 \,\widetilde{I}_{1,5,1}+4
   \,\widetilde{I}_{3,1,3}+4 \,\widetilde{I}_{3,3,1}-32 \,\widetilde{I}_{1,1,1,3,1}
+\frac{2 \,\widetilde{I}_{1,3}}{(n+1)^2}+8 \,\widetilde{I}_{1,5}+\nonumber\\ 
&& + \frac{6
   \,\widetilde{I}_{3,1}}{(n+1)^2}+4 \,\widetilde{I}_{3,3}+4 \,\widetilde{I}_{5,1}-32
   \,\widetilde{I}_{1,1,3,1}-\frac{13}{2 (n+1)^6}
+\frac{4 \,\widetilde{I}_1}{(n+1)^4}+\nonumber\\
&& + \frac{2 \,\widetilde{I}_3}{(n+1)^2}-\frac{\pi ^2
   \,\widetilde{I}_3}{12 (n+1)^2}-\frac{\pi ^2 \,\widetilde{I}_5}{4}+4
   \,\widetilde{I}_5+\frac{2}{3} \pi ^2 \,\widetilde{I}_{1,1,3}-16 \,\widetilde{I}_{1,3,1} + \nonumber\\
&& 
-2 \zeta _3 \,\widetilde{I}_{1,3}+\frac{2}{3} \pi ^2 \,\widetilde{I}_{1,3}+8
   \,\widetilde{I}_{1,3}-8 \,\widetilde{I}_{3,1}-2 \,\widetilde{I}_{3,1} \zeta _3+\nonumber\\
&& + \frac{\zeta
   _3}{(n+1)^4}+\frac{4}{(n+1)^4}-\frac{\pi ^2}{3 (n+1)^4}
-\frac{4 \,\widetilde{I}_1}{(n+1)^2}-\frac{\pi ^4 \,\widetilde{I}_1}{30
   (n+1)^2}-\frac{\pi ^4 \,\widetilde{I}_3}{15}+\frac{\pi ^2 \,\widetilde{I}_3}{3}+8
   \,\widetilde{I}_3+\nonumber\\ 
&& + \frac{4}{15} \pi ^4 \,\widetilde{I}_{1,1,1}-2 \,\widetilde{I}_3 \zeta _3
-\frac{2}{3} \pi ^2 \zeta _3 \,\widetilde{I}_{1,1}-32 \zeta _5
   \,\widetilde{I}_{1,1}+\frac{4}{15} \pi ^4 \,\widetilde{I}_{1,1}+\nonumber\\
&& + \frac{\pi ^2 \zeta
   _3}{6 (n+1)^2}+\frac{8 \zeta _5}{(n+1)^2}+\frac{24}{(n+1)^2}-\frac{\pi
   ^4}{15 (n+1)^2}+\frac{2 \pi ^2}{3 (n+1)^2}
-\frac{2}{3} \pi ^2 \zeta _3 \,\widetilde{I}_1-8 \zeta _3 \,\widetilde{I}_1+\nonumber\\
&& -32 \zeta _5
   \,\widetilde{I}_1-\frac{7 \pi ^6 \,\widetilde{I}_1}{270}+\frac{2 \pi ^4
   \,\widetilde{I}_1}{15}
-\frac{2}{3} \pi ^2 \zeta _3-16 \zeta _3-32 \zeta _5-\frac{7 \pi
   ^6}{135}-\frac{4 \pi ^4}{15}-\frac{16 \pi ^2}{3}-160.
\ea
Notice that we did not attempt to rearrange it in any minimal form.

Finally, the dressing contribution reads
\be
{\cal P}_4^{\rm dressing} = -4\,\widetilde{I}_{1,3}-4\,\widetilde{I}_{3,1}-4\,\widetilde{I}_3-4\,\frac{\widetilde{I}_1}{(n+1)^2}-\frac{8}{(n+1)^2}+\frac{2}{(n+1)^4}
\ee
and, as anticipated, is {\em separately} reciprocity respecting.

As a consequence of reciprocity, it is possible to analyze the large $N$ expansion of the four loop anomalous dimension and of ${\cal P}$
in view  of the MVV relations. This is a technical issue which is presented in Appendix~(\ref{app:expansions}).

\section{Reciprocity and wrapping}
\label{sec:wrapping}

We have presented our multi-loop result and its analysis without much worry about possible wrapping problems.
In this brief section, we make a few remarks about this important issue.

It is well-known that the long-range Bethe Ansatz equations are only asymptotic~\cite{Beisert:2005fw}. The length of the chain (and thus of the operator) 
is assumed to exceed the range of the interaction (and thus the order in perturbation theory), reaching the asymptotic conditions by which the S-matrix can be defined according to the 
perturbative Bethe Ansatz technique~\cite{Staudacher:2004tk}.

If the interaction range of the dilatation operator reaches or exceeds the length of the operator under study,  the Bethe  ansatz might 
break down~\cite{Ambjorn:2005wa,SchaferNameki:2006ey,Kotikov:2007cy}.
In special subsectors, as $\mathfrak{su}(2)$, higher order expressions of the dilatation operator are known and this issue can be checked in full details~\cite{Beisert:2004hm,Beisert:2007hz}.
In other cases, like in the $\mathfrak{sl}(2)$ sector, supersymmetry can be invoked to explain special delays of the wrapping phenomenon~\cite{Staudacher:2004tk}.

In our calculation, such tools are not (yet) available and we cannot prove nor exclude wrapping effects at 3 or 4 
loops~\footnote{The two loop case seem reasonably safe for length 3 states, in that interactions are still only between next-to-nearest neighbors.}.
What we have proved rigorously is that {\em the asymptotic Bethe Ansatz predicts a result which is reciprocity respecting}. We believe that this is an interesting 
result {\em per se}, pointing toward hidden properties of the Bethe equations. Besides, we emphasize that it would be incorrect to believe that a 
reciprocity respecting result means that wrapping effects are absent. If one believes that reciprocity is a physically meaningful property, 
it could simply be that the (yet to be quantified) wrapping-correction is also reciprocity respecting.

As a sort of example of this phenomenon we can exhibit a case where the asymptotic Bethe Ansatz provides a result which is certainly wrong, {\em i.e.}
misses the wrapping contributions, but nevertheless is reciprocity respecting. This is the four loop prediction for the Konishi operator 
reported in~\cite{Kotikov:2007cy}  and known to violate the BFKL equation as well as both of the recent (not coinciding) field theoretical calculations~\cite{Fiamberti:2007rj,Keeler:2008ce}.

For the first three loops, it has been proved that the $\cal P$ function satisfies reciprocity in all orders~\cite{Dokshitzer:2006nm}. 
In terms of a series expansion in $1/J^2$ and for the first few orders, it reads~\footnote{Notice that the notation adopted in~\cite{Kotikov:2007cy}, in which $g^2=\frac{\lambda}{16\pi^2}$, differs from the one used here. }
\ba
{\cal P}_1&=&4 \log J+\frac{2}{3}\frac{1}{J^2}-\frac{2}{15}\frac{1} {J^4}+{\cal O}\left(\frac{1}{J^{6}}\right)\,,\\
   {\cal P}_2&=&-\frac{2}{3} \pi ^2 \log J-6 \zeta_3+\left(2-\frac{\pi ^2}{9}\right)\frac{1}{J^2}+\left(1+\frac{\pi
   ^2}{45}\right)\frac{1}{J^4}+{\cal O}\left(\frac{1}{J^6}\right)\,,\\\nonumber
   {\cal P}_3&=&\frac{11}{45} \pi ^4 \log J+\frac{2}{3} \pi ^2 \zeta_3+20 \zeta_5+\left(\frac{11 \pi ^4}{270}-\frac{2}{3} \pi
   ^2 \log J\right)\frac{1}{J^2}\\&&-\left[2 +\frac{7 \pi ^2}{9}+\frac{11 \pi
   ^4}{1350}-2\left(3+\frac{\pi ^2}{3}\right) \log J\right]\frac{1}{J^4}
 +{\cal O}\left(\frac{1}{J^{6}}\right)\,.
   \ea
At four loops, starting from~\cite{Kotikov:2007cy}, we derived a series expansion for ${\cal P}_4$ that reads 
\ba\nonumber
{\cal P}_4&=&-\left (\frac {73\pi^6} {630} + 4\zeta_ 3^2 \right)\log J-\frac{7}{30} \pi ^4
   \zeta_3-\frac{5}{3} \pi ^2 \zeta_5-\frac{175}{2} \zeta_7   
   \\\nonumber
   &&-\left[\frac{\pi ^4}{30}+\frac{73 \pi ^6}{3780}-\pi ^2 \zeta_3+\frac{2 \zeta_3^2}{3}-\left(\frac{7 \pi ^4}{15}+4 \zeta_3 \right)\log J+8 \zeta_3\log^2 J\right]\frac{1}{J^2}\\\nonumber&&
+\left[\frac{1}{2}-\frac{\pi^2}{2}+\frac{71 \pi ^4}{180}+\frac{73 \pi ^6}{18900}-\left(\pi^2+\frac{25}{3}\right)\zeta_3+\frac{2 \zeta_3^2}{15}-\left(\pi ^2+\frac{7 \pi ^4}{15}+\frac{26 \zeta_3}{3}\right)\log J\right.\\&&\left.
+4 \left(\frac{\pi ^2}{3}+2 \zeta_3 \right) \log^2 J+8 \log^3 J\right]\frac{1}{J^4}
+{\cal O}\left(\frac{1}{J^{6}}\right)
\ea   Only integer negative powers of $J^2$ appear (even extending the series by many orders) proving (empirically) that reciprocity holds.

\section{Conclusions}

We have considered a special class of scaling composite operators in ${\cal N}=4$ SYM which at one-loop admits a simple description as gluonic
quasipartonic twist operators. We have been able to compute their anomalous dimension at 4 loops in the framework of the asymptotic long-range 
Bethe Ansatz. This has been possible by formulating a suitable generalized transcendentality principle leading to an inspired Ansatz 
in terms of nested harmonic sums. The main test of our result has been to show that it respects the generalized Gribov-Lipatov reciprocity 
recently discovered in other sectors for the Dokshizter-Marchesini-Salam evolution kernel.

Going back to our initial motivations, we see that the large spin analysis of twist operators is indeed rich and  somewhat surprising. 
The general structure of the expansion has a well understood leading logarithmic term which can be resummed in terms of the physical coupling 
governing soft radiation effects. The physical coupling must emerge in a universal way in all conformal sectors (scalars, gauginos or gauge fields) 
and, presumably, for all twists (with positive checks in the $L=2, 3, \infty$ cases). 
The mechanism is also clear on the AdS side, as explained for instance in~\cite{Belitsky:2003ys} and in the recent 
analysis~\cite{Kruczenski:2004wg,Frolov:2006qe,Kruczenski:2008bs}, although a better identification of the string solution dual to 
the {\em minimal} gluonic operator would be welcome. Indeed, It is known that anomalous 
dimensions of operators with twist higher than two occupy a band~\cite{Belitsky:2003ys}, whose lower bound is the one of interest
 in this paper. The spiky strings proposed in~\cite{Kruczenski:2004wg} are dual to higher twist operators with maximal anomalous dimension.
In addition, a more general problem of identification follows from the fact that the field strength does not carry $R$-charge. While 
it is natural to guess that operators built out of many covariant derivatives and field strength components should correspond to 
strings stretched in AdS having large spin, it is not clear  how one could distinguish between scalars, fermions or the field 
strength without  the guidance from some {\em extra} charge easily visible on both sides of the AdS/CFT.

On the other hand, the constraints on the subleading terms at large $N$ implied by reciprocity have a much less clear origin.
In particular, it seems that a general reciprocity proof is missing in the gauge theory. 
Indeed, we have found empirically that reciprocity holds in
many cases with various conformal spins and twists, but the details of the derivation are drastically non-universal.
The reason is that the reciprocity proofs heavily rely on the detailed (closed) form of the spin dependent anomalous dimensions.
Unfortunately, we miss a unifying principle treating uniformly the various known cases.
Also, what is the dual counterpart of reciprocity? In~\cite{Basso:2006nk}, reciprocity is tested at strong coupling for the
semiclassical string configuration dual to the minimal anomalous dimension $\mathfrak{sl}(2)$ twist-$L$ operator.
This is the folded string rotating with angular momentum $N$ on $\mbox{AdS}_3$ and with center of mass moving with angular 
momentum $L$ on a big circle of $S^5$~\cite{Gubser:2002tv,Frolov:2002av}. 
An extension to string states dual to 
other reciprocity respecting gauge theory operators would certainly be welcome.

As a final comment, we emphasize that the observed four loop reciprocity for gauge operators 
must still pass the test of wrapping effects, as discussed in Sec.~(\ref{sec:wrapping}). 
Nevertheless, it certainly suggests some important structure built in the Bethe Ansatz and deserving a deeper understanding.
As we learn from the twist-2 QCD lesson, the attempts to extend at higher loop orders the Gribov-Lipatov relation 
led to the discovery of the DMS reciprocity respecting kernel. This innovative rewriting of parton evolution revealed new 
relations between space and time-like anomalous dimensions. In perspective, we believe that 
the observation of an intrinsic reciprocity in the {\em asymptotic} Bethe Ansatz equations of ${\cal N}=4$ SYM should not be regarded as a mere technical feature. Instead, it could be a starting point to constrain the still elusive {\em wrapping} corrections.

\acknowledgments

We thank G. Marchesini, Y. L. Dokshitzer, G. Korchemsky, A. Belitski, J. Plefka,  A. Tseytlin and in particular M. Staudacher for discussions.
M.~B. warmly thanks the Physics Department of Humboldt University Berlin and the Albert Einstein Institute in Golm, Potsdam for the very kind hospitality while working on parts of this project. 

\appendix

\section{Large $N$ expansions and reciprocity}
\label{app:expansions}

>From reciprocity of ${\cal P}$ one immediately proves that (basically) half of the terms in $\gamma$ are truly independent
in the spirit of the MVV constraints.
This is discussed in~\cite{Basso:2006nk} that we now follow. The relation between $\gamma$ and ${\cal P}$ can be formally solved  by means of the 
Lagrange-B\"urmann formula leading to 
\ba
\gamma(N) &=& \sum_{k=1}^\infty \frac{1}{k!}\left(\frac{1}{2}\,\frac{\partial}{\partial N}\right)^{k-1}\,[{\cal P}(N)]^k, \\
{\cal P}(N) &=& \sum_{k=1}^\infty \frac{1}{k!}\left(-\frac{1}{2}\,\frac{\partial}{\partial N}\right)^{k-1}\,[\gamma(N)]^k.
\ea
If we separate 
\be
\gamma(N) = \gamma_+(N) + \gamma_-(N),
\ee
with
\ba
\gamma_+(N) &=& \sum_{k=0}^\infty \frac{1}{(2\,k+1)!}\left(\frac{1}{2}\,\partial\right)^{2\,k} [{\cal P}(N)]^{2\,k+1}, \\
\gamma_-(N) &=& \sum_{k=1}^\infty \frac{1}{(2\,k)!}\left(\frac{1}{2}\,\partial\right)^{2\,k-1} [{\cal P}(N)]^{2\,k},
\ea
then it can be shown that a reciprocity respecting kernel leads to the constraint
\be
\label{eq:constraint}
\gamma_- = \frac{1}{4} (\gamma_+^2)' + \frac{1}{48}\left(-\gamma_+ (\gamma_+^3)''+\frac{1}{4} (\gamma_+^4)''\right)' + \cdots.
\ee
Expanding in loops, we find
\ba
\gamma_{-,1} &=& 0, \\
\gamma_{-,2} &=& \frac{1}{4} (\gamma_{+,1}^2)', \\
\gamma_{-,3} &=& \frac{1}{2} (\gamma_{+,1}\,\gamma_{+,2})', \\
\gamma_{-,4} &=& \frac{1}{4}(\gamma_{+,2}^2+2\,\gamma_{+,1}\,\gamma_{+,3})' + \frac{1}{48}\left(-\gamma_{+,1} (\gamma_{+,1}^3)''
+\frac{1}{4} (\gamma_{+,1}^4)''\right)' .
\ea
However, these relations are of little practical use. They are completely equivalent to MVV relations that are more
transparent since directly  connect specific terms in the large $N$ expansion of $\gamma$. To this aim it is convenient to rewrite the 
most difficult $\gamma_4$ piece in terms of $S_1$ and harmonic sums which are convergent as $N\to\infty$.

\subsection{Reversed form of $\gamma_4$, suitable for the large $N$ expansion}

Using the shuffle algebra we rewrite the maximal \tran term in $\gamma_4^{\rm no\ dressing}$ as 

\ba
{\cal H}_{7,7} &=&
\frac{40}{3} S_4 S_1^3-\frac{32}{3} S_{3,1} S_1^3+20 S_5 S_1^2-40 S_{3,2}
   S_1^2-56 S_{4,1} S_1^2+64 S_{3,1,1} S_1^2-4 S_2^3 S_1+32 S_3^2 S_1 + \nonumber\\
&& -4 S_2
   S_4 S_1+39 S_6 S_1+88 S_2 S_{3,1} S_1-64 S_{4,2} S_1+48 S_{5,1} S_1-104
   S_{2,3,1} S_1+120 S_{4,1,1} S_1 + \nonumber\\
&& -192 S_{3,1,1,1} S_1-2 S_2^2 S_3-\frac{289
   S_3 S_4}{3}-S_2 S_5-\frac{189 S_7}{2}-\frac{256}{3} S_3 S_{3,1}-4 S_2
   S_{3,2}+60 S_2 S_{4,1} + \nonumber \\
&& +136 S_{4,3}-24 S_{5,2}-32 S_{6,1}-64 S_{2,4,1}-120
   S_2 S_{3,1,1}+64 S_{3,2,2}+80 S_{3,3,1}-80 S_{5,1,1} + \nonumber \\
&& +128 S_{2,3,1,1}-128
   S_{4,1,1,1}+256 S_{3,1,1,1,1}, \nonumber\\
{\cal H}_{7,6} &=&
-2 S_2^3-2 S_4 S_2+44 S_{3,1} S_2+16 S_3^2+20 S_1^2 S_4+20 S_1 S_5+\frac{39
   S_6}{2}-16 S_1^2 S_{3,1}-40 S_1 S_{3,2} + \nonumber\\
&& -56 S_1 S_{4,1}-32 S_{4,2}+24
   S_{5,1}-52 S_{2,3,1}+64 S_1 S_{3,1,1}+60 S_{4,1,1}-96 S_{3,1,1,1}, \nonumber\\
{\cal H}_{7,5} &=&
-6 S_1 S_2^2-2 S_3 S_2+8 S_1 S_4+\frac{9 S_5}{2}-16 S_1 S_{3,1}-12 S_{3,2}-16
   S_{4,1}+20 S_{3,1,1}, \nonumber\\
{\cal H}_{7,4} &=&
-6 S_2 S_1^2-4 S_2^2+\frac{S_4}{2}-8 S_{3,1}, \nonumber\\
{\cal H}_{7,3} &=&
-2 S_1^3-15 S_2 S_1-2 S_3, \nonumber\\
{\cal H}_{7,2} &=&
-12 S_1^2-8 S_2 ,\nonumber\\
{\cal H}_{7,1} &=&
-\frac{39 S_1}{2}, \nonumber\\
{\cal H}_{7,0} &=&
-\frac{39}{4} .
\ea
The other pieces of $\gamma_4^{\rm no \ dressing}$ are

\ba
{\cal H}_{6,6} &=&
-4 S_2^3-4 S_4 S_2+88 S_{3,1} S_2+32 S_3^2+40 S_1^2 S_4+40 S_1 S_5+39 S_6-32
   S_1^2 S_{3,1}-80 S_1 S_{3,2} + \nonumber\\
&& -112 S_1 S_{4,1}-64 S_{4,2}+48 S_{5,1}-104
   S_{2,3,1}+128 S_1 S_{3,1,1}+120 S_{4,1,1}-192 S_{3,1,1,1}, \nonumber\\
{\cal H}_{6,5} &=&
40 S_1 S_4+20 S_5-32 S_1 S_{3,1}-40 S_{3,2}-56 S_{4,1}+64 S_{3,1,1}, \nonumber\\
{\cal H}_{6,4} &=&
-6 S_2^2+4 S_1 S_3+8 S_4-16 S_{3,1} ,\nonumber\\
{\cal H}_{6,3} &=&
2 S_3-12 S_1 S_2, \nonumber\\
{\cal H}_{6,2} &=&
-6 S_1^2-15 S_2, \nonumber\\
{\cal H}_{6,1} &=&
-22 S_1, \nonumber\\
{\cal H}_{6,0} &=&
-\frac{37}{2}, \nonumber \\
{\cal H}_{5,5} &=&
40 S_1 S_4+20 S_5-32 S_1 S_{3,1}-40 S_{3,2}-56 S_{4,1}+64 S_{3,1,1}, \nonumber\\
{\cal H}_{5,4} &=&
20 S_4-16 S_{3,1}, \nonumber\\
{\cal H}_{5,3} &=&
4 S_3, \nonumber\\
{\cal H}_{5,2} &=&
-4 S_2, \nonumber\\
{\cal H}_{5,1} &=&
0, \nonumber\\
{\cal H}_{5,0} &=&
-2, \nonumber\\
{\cal H}_{4,4} &=&
-8 S_2^2+12 S_4-32 S_{3,1}, \nonumber\\
{\cal H}_{4,3} &=&
0, \nonumber\\
{\cal H}_{4,2} &=&
-4 S_2, \nonumber\\
{\cal H}_{4,1} &=&
-8 S_1, \nonumber\\
{\cal H}_{4,0} &=&
2, \nonumber \\
{\cal H}_{3,1} &=&
-8 S_1, \nonumber\\
{\cal H}_{3,0} &=&
4, \nonumber \\
{\cal H}_{2,2} &=&
-32 S_2, \nonumber\\
{\cal H}_{2,1} &=&
0, \nonumber\\
{\cal H}_{2,0} &=&
8, \nonumber \\
{\cal H}_{0,0} &=& 
-160 .
\ea

\subsection{Large $N$ expansion of $\gamma$}
\label{app:gammaexpansion}

Starting from this form of $\gamma_4$ we can easily compute the large $N$ expansion by expanding each nested sum starting from the most 
inner index. The procedure is described in full details in the Appendix of \cite{Beccaria:2007cn}. We always write the results in terms of 
$n=\frac{N}{2}+1$ and also define $\overline{n} = n\,e^{\gamma_E}$. The one-loop result is 
\be
\gamma_1 = (4 \log\,\overline{n}+4)+\frac{4}{n}-\frac{7}{3} \left(\frac{1}{n}\right)^2+2 \left(\frac{1}{n}\right)^3-\frac{59}{30} \left(\frac{1}{n}\right)^4+2
   \left(\frac{1}{n}\right)^5-\frac{127}{63} \left(\frac{1}{n}\right)^6+\cdots .
\ee
The two loop results has single logarithms in all terms 
\ba
\gamma_2 &=&
\left(-\frac{2 \pi ^2 \log\,\overline{n}}{3}-2 \zeta _3-\frac{2 \pi ^2}{3}-8\right)+\frac{4 \log\,\overline{n}-\frac{2 \pi ^2}{3}+4}{n}+\left(-4 \log\,\overline{n}+\frac{7 \pi
   ^2}{18}+1\right) \left(\frac{1}{n}\right)^2+\nonumber\\
&& + \left(\frac{14 \log\,\overline{n}}{3}-\frac{\pi ^2}{3}-\frac{11}{3}\right) \left(\frac{1}{n}\right)^3+\left(-6
   \log\,\overline{n}+\frac{59 \pi ^2}{180}+\frac{13}{2}\right) \left(\frac{1}{n}\right)^4+\nonumber\\
&& + \left(\frac{118 \log\,\overline{n}}{15}-\frac{\pi ^2}{3}-\frac{487}{45}\right)
   \left(\frac{1}{n}\right)^5+\left(-10 \log\,\overline{n}+\frac{127 \pi ^2}{378}+\frac{35}{2}\right)
   \left(\frac{1}{n}\right)^6+\cdots\ .
\ea
At three loops, we find quadratic logarithms starting from the $1/n^2$ term 
\ba
\gamma_3 &=& 
\left(\frac{11 \pi ^4 \log\,\overline{n}}{45}+\frac{\pi ^2 \zeta _3}{3}-\zeta _5+\frac{11 \pi ^4}{45}+\frac{4 \pi ^2}{3}+32\right)+\frac{-\frac{4 \pi ^2
   \log\,\overline{n}}{3}-2 \zeta _3+\frac{11 \pi ^4}{45}-\frac{4 \pi ^2}{3}-8}{n}+\nonumber\\
&& + \left(-2 \log^2\,\overline{n}+\frac{4 \pi ^2 \log\,\overline{n}}{3}+2 \zeta _3-\frac{77 
\pi^4}{540}-\frac{\pi ^2}{6}+7\right) \left(\frac{1}{n}\right)^2+\\
&& + \left(4 \log^2\,\overline{n}+\left(-6-\frac{14 \pi ^2}{9}\right) \log\,\overline{n}-\frac{7 
\zeta_3}{3}+\frac{11 \pi ^4}{90}+\frac{8 \pi ^2}{9}-\frac{25}{3}\right) \left(\frac{1}{n}\right)^3+\nonumber\\
&& + \left(-7 \log^2\,\overline{n}+\left(\frac{47}{3}+2 \pi ^2\right)
   \log\,\overline{n}+3 \zeta _3-\frac{649 \pi ^4}{5400}-\frac{19 \pi ^2}{12}+\frac{227}{24}\right) \left(\frac{1}{n}\right)^4+\nonumber\\
&& + \left(12
   \log^2\,\overline{n}+\left(-32-\frac{118 \pi ^2}{45}\right) \log\,\overline{n}-\frac{59 \zeta _3}{15}+\frac{11 \pi ^4}{90}+\frac{352 \pi ^2}{135}-\frac{181}{15}\right)
   \left(\frac{1}{n}\right)^5+\nonumber\\
&& + \left(-\frac{59 \log^2\,\overline{n}}{3}+\left(\frac{2789}{45}+\frac{10 \pi ^2}{3}\right) \log\,\overline{n}+5 \zeta _3-\frac{1397 
\pi^4}{11340}-\frac{151 \pi ^2}{36}+\frac{4033}{270}\right) \left(\frac{1}{n}\right)^6+\cdots .\nonumber
\ea
The non-dressing four loops anomalous dimension has a quadratic logarithm in the $1/n^2$ term and cubic logarithms in all the subsequent ones. It reads
\ba
\gamma_4^{\rm no \ dressing} &=& 
\left(4 \zeta _3^2-\frac{2 \pi ^4 \zeta _3}{15}+\log\,\overline{n} \left(4 \zeta _3^2-\frac{73 \pi ^6}{630}\right)+\frac{\pi ^2 \zeta _5}{6}+\frac{55 \zeta
   _7}{2}-\frac{73 \pi ^6}{630}-\frac{8 \pi ^4}{15}-\frac{16 \pi ^2}{3}-160\right)+\nonumber\\
&& + \frac{4 \zeta _3^2+\frac{2 \pi ^2 \zeta _3}{3}-\zeta _5-\frac{73 \pi
   ^6}{630}+\frac{3 \log\,\overline{n} \pi ^4}{5}+\frac{3 \pi ^4}{5}+\frac{8 \pi ^2}{3}+32}{n}+\nonumber\\
&& + \left(\pi ^2 \log^2\,\overline{n}+\left(4 \zeta _3-\frac{3 \pi
   ^4}{5}+4\right) \log\,\overline{n}-\frac{7 \zeta _3^2}{3}-\frac{2 \pi ^2 \zeta _3}{3}+2 \zeta _3+\zeta _5\right. \nonumber\\
&& \left. +\frac{73 \pi ^6}{1080}+\frac{\pi ^4}{15}-\frac{19 \pi
   ^2}{6}-12\right) \left(\frac{1}{n}\right)^2+\left(\frac{4 \log^3\,\overline{n}}{3}+\left(-2-2 \pi ^2\right) \log^2\,\overline{n}\right. \nonumber\\
&& \left. +\left(-8 \zeta _3+\frac{7 \pi
   ^4}{10}+\frac{8 \pi ^2}{3}-6\right) \log\,\overline{n}+2 \zeta _3^2+\frac{7 \pi ^2 \zeta _3}{9}+\zeta _3-\frac{7 \zeta _5}{6}-\frac{73 \pi ^6}{1260} \right.\nonumber\\
&& \left. -\frac{23 \pi
   ^4}{60}+\frac{40 \pi ^2}{9}+\frac{23}{3}\right) \left(\frac{1}{n}\right)^3+\left(-4 \log^3\,\overline{n}+\left(13+\frac{7 \pi ^2}{2}\right)
   \log^2\,\overline{n}\right.\nonumber\\
&& \left. +\left(12 \zeta _3-\frac{9 \pi ^4}{10}-\frac{41 \pi ^2}{6}+\frac{47}{4}\right) \log\,\overline{n}-\frac{59 \zeta _3^2}{30}-\pi ^2 \zeta _3-8 \zeta
   _3+\frac{3 \zeta _5}{2}+\frac{4307 \pi ^6}{75600}+\right.\nonumber\\
&& \left. \frac{41 \pi ^4}{60}-\frac{83 \pi ^2}{16}-\frac{65}{4}\right)
   \left(\frac{1}{n}\right)^4+\cdots \ .
\ea
Finally, the dressing part has the expansion
\ba
\gamma_4^{\rm dressing} &=& 
\left(-8 \log\,\overline{n} \zeta _3-8 \zeta _3\right)-\frac{8 \zeta _3}{n}+\left(-4 \log\,\overline{n}+\frac{14 \zeta _3}{3}-4\right) \left(\frac{1}{n}\right)^2+\left(8
   \log\,\overline{n}-4 \zeta _3+4\right) \left(\frac{1}{n}\right)^3+\nonumber\\
&& \left(-10 \log\,\overline{n}+\frac{59 \zeta _3}{15}+\frac{1}{3}\right)
   \left(\frac{1}{n}\right)^4+\cdots\ .
\ea

\subsection{MVV-like relations}

For simplicity we set $\gamma_E\to 0$,
which does no loose information since all logarithms have as a natural argument the combination $\overline n=n^{\gamma_E}$. 
We write the general expansion of $\gamma$ as
\ba
\gamma(n) &=& L_{0,1}\,\log\,n + c_0 + \frac{L_{1,1}\,\log\,n+c_1}{n} + \frac{L_{2,2}\,\log^2\,n+L_{2,1}\,\log\,n+c_2}{n^2}+\nonumber \\
&& + \frac{L_{3,3}\,\log^3\,n + L_{3,2}\,\log^2\,n+L_{3,1}\,\log\,n+c_3}{n^3}+{\cal O}\left(\frac{\log^3\,n}{n^4}\right),
\ea
where $L_{ij}$ and $c_i$ are functions of the coupling.

The most general expansion of a reciprocity respecting ${\cal P}(N)$ compatible with the large $N$ expansion of $\gamma$ is
\be
{\cal P}(N) = p_{0,1}\,\log\frac{N(N+8)}{4}+b_0+\frac{p_{1,2}\,\log^2\frac{N(N+8)}{4}+p_{1,1}\,\log\frac{N(N+8)}{4}+b_1}{N(N+8)} + {\cal O}\left(
\frac{\log^3 N}{N^4}\right).
\ee
Matching the above two expansions in the relation 
\be
\gamma(n) = {\cal P}\left(N+\frac{1}{2}\,\gamma(n)\right),\qquad n = \frac{N}{2}+1,
\ee
we determine all the coefficients in the expansion of ${\cal P}$ and also find a set of constraints on the coefficients of the expansion of $\gamma$. These 
constraints give all the terms of the form $(\log\,n)^p/n^{2\,q+1}$ in terms of those of the form $(\log\,n)^p/n^{2\,q}$. The precise relations are 
the following lowest order MVV relations
\ba
L_{1,1} &=& \frac{L_{0,1}^2}{4}, \\
c_1 &=& \frac{1}{4} c_0 L_{0,1}+L_{0,1}, 
\ea
and the successive ones
\ba
L_{3,3} &=&  -\frac{1}{96} \,L_{0,1}^4-\frac{1}{2} \,L_{2,2} \,L_{0,1}, \\
L_{3,2} &=& \frac{L_{0,1}^4}{32}-\frac{1}{32} \,c_0 \,L_{0,1}^3-\frac{L_{0,1}^3}{8}-\frac{1}{2} \,L_{2,1} \,L_{0,1}+\frac{3}{4} \,L_{2,2} \,L_{0,1}-\frac{1}{2} \,c_0 \,L_{2,2}-2
   \,L_{2,2}, \\
L_{3,1} &=& -\frac{1}{64} \,L_{0,1}^4+\frac{1}{16} \,c_0 \,L_{0,1}^3+\frac{L_{0,1}^3}{4}-\frac{1}{32} \,c_0^2 \,L_{0,1}^2-\frac{1}{4} \,c_0
   \,L_{0,1}^2-\frac{L_{0,1}^2}{2}-\frac{1}{2} \,c_2 \,L_{0,1}+\frac{1}{2} \,L_{2,1} \,L_{0,1} \nonumber\\
&& -\frac{1}{2} \,c_0 \,L_{2,1}-2 \,L_{2,1}+\frac{1}{2} \,c_0 \,L_{2,2}+2 \,L_{2,2}, \\
c_3 &=&
   -\frac{1}{96} \,L_{0,1} \,c_0^3+\frac{1}{32} \,L_{0,1}^2 \,c_0^2-\frac{1}{8} L_{0,1} \,c_0^2-\frac{1}{64} \,L_{0,1}^3 \,c_0+\frac{1}{4} \,L_{0,1}^2 \,c_0-\frac{c_2
   \,c_0}{2}-\frac{1}{2} \,L_{0,1} \,c_0+\frac{1}{4} \,L_{2,1} \,c_0 \nonumber\\
&& -\frac{L_{0,1}^3}{16}+\frac{L_{0,1}^2}{2}-2 \,c_2+\frac{1}{4} \,c_2 \,L_{0,1}-\frac{2
   \,L_{0,1}}{3}+\,L_{2,1}
\ea
The explicit values of these coefficients for the canonical choice $\beta=\zeta_3$, {\em i.e.}
\be
\gamma_4 = \gamma_4^{\rm no \ dressing} + \zeta_3\,\gamma_4^{\rm dressing},
\ee
are 
\ba
L_{0,1} &=& 4 g^2-\frac{2 \pi ^2 g^4}{3}+\frac{11 \pi ^4 g^6}{45}+\left(-4 \zeta _3^2-\frac{73 \pi ^6}{630}\right) g^8+\cdots , \\
c_0 &=& 4 g^2+\left(-2 \zeta _3-\frac{2 \pi ^2}{3}-8\right) g^4+\left(\frac{\pi ^2 \zeta _3}{3}-\zeta _5+\frac{11 \pi ^4}{45}+\frac{4 \pi ^2}{3}+32\right)
   g^6+\cdots ,\\
L_{1,1} &=& 4 g^4-\frac{4 \pi ^2 g^6}{3}+\frac{3 \pi ^4 g^8}{5}+\cdots , \\
c_1 &=& 4 g^2+\left(4-\frac{2 \pi ^2}{3}\right) g^4+\left(-2 \zeta _3+\frac{11 \pi ^4}{45}-\frac{4 \pi ^2}{3}-8\right) g^6+\nonumber\\
&& + \left(-4 \zeta _3^2+\frac{2 \pi ^2 \zeta
   _3}{3}-\zeta _5-\frac{73 \pi ^6}{630}+\frac{3 \pi ^4}{5}+\frac{8 \pi ^2}{3}+32\right) g^8+\cdots , \\
L_{2,2} &=& -2 g^6+\pi ^2 g^8+\cdots , \\
L_{2,1} &=& -4 g^4+\frac{4 \pi ^2 g^6}{3}+\left(4-\frac{3 \pi ^4}{5}\right) g^8+\cdots , \\
c_2 &=& -\frac{7 g^2}{3}+\left(1+\frac{7 \pi ^2}{18}\right) g^4+\left(2 \zeta _3-\frac{77 \pi ^4}{540}-\frac{\pi ^2}{6}+7\right) g^6+\nonumber\\
&& + \left(\frac{7 \zeta _3^2}{3}-\frac{2
   \pi ^2 \zeta _3}{3}-2 \zeta _3+\zeta _5+\frac{73 \pi ^6}{1080}+\frac{\pi ^4}{15}-\frac{19 \pi ^2}{6}-12\right) g^8+\cdots , \\
L_{3,3} &=& \frac{4 g^8}{3}+\cdots , \\
L_{3,2} &=& 4 g^6+\left(-2-2 \pi ^2\right) g^8+\cdots , \\
L_{3,1} &=& \frac{14 g^4}{3}+\left(-6-\frac{14 \pi ^2}{9}\right) g^6+\left(-6+\frac{8 \pi ^2}{3}+\frac{7 \pi ^4}{10}\right) g^8+\cdots , \\
c_3 &=& 2 g^2+\left(-\frac{11}{3}-\frac{\pi ^2}{3}\right) g^4+\left(-\frac{7 \zeta _3}{3}+\frac{11 \pi ^4}{90}+\frac{8 \pi ^2}{9}-\frac{25}{3}\right) g^6+\nonumber\\
&& + \left(-2 \zeta
   _3^2+\frac{7 \pi ^2 \zeta _3}{9}+5 \zeta _3-\frac{7 \zeta _5}{6}-\frac{73 \pi ^6}{1260}-\frac{23 \pi ^4}{60}+\frac{40 \pi ^2}{9}+\frac{23}{3}\right)
   g^8+\cdots 
\ea
Notice that the four loop contribution to $c_0$ does not enter the above relations but only higher order ones. Also, 
the relations are true irrespectively on $\beta$ since the dressing part is separately reciprocity respecting.

It is a straightforward exercise to check that these expressions indeed obey the MVV relations.

\subsection{Large $N$ expansion of ${\cal P}$}

In the spirit of the analysis of \cite{Dokshitzer:2006nm} and \cite{Beccaria:2007bb} we present the large $N$ expansion of ${\cal P}$
once it is re-expanded in terms of the physical coupling $g^2_{\rm ph} = \frac{1}{2}\,\Gamma_{\rm cusp}$ which reads at 4 loops
\ba
\gamma &=& 4\,g^2_{\rm ph}\,\log\,N + {\cal O}(N^0), \\
g^2_{\rm ph} &=& g^2-\frac{\pi^2}{6}\,g^4+\frac{11\,\pi^4}{180}\,g^6-\frac{1}{4}\left(\frac{73\,\pi^6}{630}+4\,\zeta_3^2\right)\,g^8+\cdots~.
\ea
The reciprocity respecting kernel ${\cal P}$ can be re-expanded in the physical coupling 
\be
{\cal P} = \sum_{n=1}^\infty {\cal P}_n^{\rm ph}\,g_{\rm ph}^{2\,n}.
\ee
The large $n$ expansion at four loops reads
\ba
{\cal P}_1^{\rm ph}(n) &=& 4\,\log\,\overline{n} + 4 + \frac{4}{n}-\frac{7}{3}\,\frac{1}{n^2}+\frac{2}{n^3}-\frac{59}{30}\,\frac{1}{n^4}+\frac{2}{n^5}+\cdots~, \\
{\cal P}_2^{\rm ph}(n) &=& -8-2\,\zeta_3+\frac{1}{n^2}-\frac{2}{n^3}+\frac{7}{2}\,\frac{1}{n^4}-\frac{6}{n^5} + \cdots~, \\
{\cal P}_3^{\rm ph}(n) &=& 32-\frac{4\,\pi^2}{4}-\frac{\pi^2}{3}\,\zeta_3-\zeta_5 + \left(\frac{\pi^2}{6}-3\right)\,\frac{1}{n^2} + 
\left(6-\frac{\pi^2}{3}\right)\,\frac{1}{n^3} + \nonumber\\
&&  +\left(-\frac{63}{8}+\frac{7\,\pi^2}{12}\right)\,\frac{1}{n^4} + \left(\frac{15}{2}-\pi^2\right)\,\frac{1}{n^5} + \cdots~, \\
{\cal P}_4^{\rm ph}(n) &=& -160+\frac{32\,\pi^2}{3}-\frac{\pi^2}{3}\,\zeta_5+\frac{55}{2}\,\zeta_7 + (20-\pi^2-2\,\zeta_3-2\,(2+\zeta_3)\,\log\,\overline{n})\,
\frac{1}{n^2} + \nonumber\\
&& + (-44+2\,\pi^2+2\,\zeta_3+4\,(2+\zeta_3)\,\log\,\overline{n})\,\frac{1}{n^3} + \cdots~.
\ea
Hence, we see that the large logarithmic terms are all hidden in the one-loop {\em physical} kernel. The next logarithmic enhancement is 
down by two powers of $n$ and starts at four loops. This is in nice agreement with what is found in the twist-3 scalar operators
analyzed in~\cite{Beccaria:2007bb} .

\section{Some technical remarks concerning harmonic sums}
\label{app:harmonic}

We collect in this Appendix some useful properties of (nested) Harmonic sums that we have used in this paper. 
Very useful references are \cite{Blumlein:2003gb}.

\subsection{Definition}

The basic definition of nested harmonic sums with positive indices $S_{a_1, \dots, a_n}$ is recursive
\ba
S_a(N) &=& \sum_{n=1}^N\frac{1}{n^a},\\
S_{a, \mathbf{b}}(N) &=& \sum_{n=1}^N\frac{1}{n^a}\, S_{\mathbf b}(n).
\ea
Given a particular sum $S_\mathbf{a} = S_{a_1, \dots, a_n}$ we define
\ba
\mbox{depth}\ (S_\mathbf{a}) &=& n, \\
\mbox{\tran} (S_\mathbf{a}) &=& |\mathbf{a}| \equiv a_1 + \cdots + a_n.
\ea
For a product of $S$ sums, we define transcendentality to be the sum of the transcendentalities of the factors.

\subsection{Shuffle algebra and canonical basis}

The basic shuffle algebra relation is 

\ba
S_a\,S_{b_1,\dots, b_k} &=& S_{a, b_1, \dots, b_k} + S_{b_1, a, b_2, \dots, b_k} + \cdots + 
S_{b_1, \dots, b_k, a} \\
&& -S_{a+b_1, \dots, b_k}-S_{b_1, a+b_2, \dots, b_k} -\cdots -S_{b_1, \dots, a+b_k}.\nonumber
\ea
It conserves the total \tran. A very useful special case is 
\be
S_a\,S_b = S_{ab} + S_{ba}-S_{a+b}.
\ee
Applying it iteratively we can reduce sums of the form $S_{a\cdots a}$ to products of simple sums of depth 1. 
In particular, we list
\ba
S_{aa} &=& \frac{1}{2}(S_a^2+S_{2a}), \\
S_{aaa} &=& \frac{1}{6}(S_a^3+3\,S_a\,S_{2a}+2\,S_{3a}), \\
S_{aaaa} &=& \frac{1}{24}(S_a^4+6\,S_a^2\,S_{2a}+3\,S_{2a}^2+8\,S_a\,S_{3a}+6\,S_{4a}).
\ea
A more general shuffle relation is 
\ba
\label{eq:shuffle}
S_{a_1, \dots, a_n}(N)\,S_{b_1, \dots, b_m}(N) &=& \sum_{\ell=1}^N \frac{1}{\ell^{a_1}}\,S_{a_2, \dots, a_n}(\ell)\,S_{b_1, \dots, b_m}(\ell) + \\
&& + \sum_{\ell=1}^N \frac{1}{\ell^{b_1}}\,S_{a_1, \dots, a_n}(\ell)\,S_{b_2, \dots, b_m}(\ell) + \nonumber \\
&& - \sum_{\ell=1}^N \frac{1}{\ell^{a_1+b_1}}\,S_{a_2, \dots, a_n}(\ell)\,S_{b_2, \dots, b_m}(\ell). \nonumber
\ea
One can apply the basic shuffle relation iteratively and prove that any product of $S$ sums can be written as a linear
combination of $S$ sums with the same total \tran .

Thus, a basis of fixed \tran $\tau$ products of sums can be reduced to single sums with varying depth. The number of such sums
can be shown to be $2^{\tau-1}$.

The first cases are $\tau=1$ with the single sum $S_1$, $\tau=2$ with the sums
\be
S_{2}, S_{11},
\ee
and $\tau=3$ with the sums
\be
S_{3}, S_{12}, S_{21}, S_{111}.
\ee
Of course, the shuffle algebra can be exploited to reduce the number of independent sums as well as to permute partially the index sets. This is useful to
isolate the large $N$ singularities of sums like $S_{1, \dots, 1, \mathbf{a}}$ in terms which are powers of $S_1$.

\subsection{Asymptotic values}

Often, it is necessary to compute $S_\mathbf{a}(\infty)$ which exists if $a_1>1$. To this aim, we define
\be
H_\mathbf{a}(N) = \sum_{N\ge n_1 > n_2 > \cdots > n_r > 0}\frac{1}{n_1^{a_1}\cdots n_r^{a_r}}.
\ee
The values of $H$ at $N=\infty$ are the so-called multiple $\zeta$ values
\be
H_\mathbf{a}(\infty) \equiv \zeta_\mathbf{a}.
\ee
The multiple zeta values are known to a large extent and are tabulated as exact combinations of elementary $\zeta$ functions.
The relation between them and $S_\mathbf{a}(\infty)$ is simple from the definition. The first cases at depth 1, 2, 3 are
\ba
S_a(\infty) &=& \zeta_a, \\
S_{a, b}(\infty) &=& \zeta_{a, b} + \zeta_{a+b}, \\
S_{a, b, c}(\infty) &=& \zeta_{a, b, c} + \zeta_{a+b, c} + \zeta_{a, b+c} + \zeta_{a+b+c}.
\ea 
The general case is obtained by summing over all possible $\zeta_\mathbf{a}$ obtained by splitting
the multiindex of $S$ in order-respecting groups ({\em i.e.} taking partitions) and taking the sum within each group. For instance
\ba
S_{a,b,c,d}(\infty) &=& 
\zeta_{a,b,c,d} + 
\zeta_{a+b,c,d} + 
\zeta_{a,b+c,d} + 
\zeta_{a,b,c+d} + 
\zeta_{a+b,c+d} + \nonumber \\
&& +
\zeta_{a+b+c,d} + 
\zeta_{a,b+c+d} + 
\zeta_{a+b+c+d}
\ea

\subsection{Derivatives}
\label{sec:der}
The analytic continuation of $S_\mathbf{a}(N)$ can be obtained from 
\be
S_{a, \mathbf{b}}(N) = \sum_{n=1}^\infty \left[\frac{1}{n^a}\,S_\mathbf{b}(n)-\frac{1}{(n+N)^a}\,S_\mathbf{b}(n+N)\right],
\ee
which can be differentiated with respect to $N$. This can be used to take derivatives of $S$ sums. 

An equivalent practical method starts from 
\be
S_{a, \mathbf{b}}(N+1)-S_{a, \mathbf{b}}(N) = \frac{1}{(N+1)^a}\,S_\mathbf{b}(N+1).
\ee
Taking a derivative and summing we find
\be
S_{a, \mathbf{b}}'(N) = -a\,S_{a+1, \mathbf{b}} + \sum_{n=1}^N \frac{1}{n^a}\,S_\mathbf{b}'(n) + c_{a, \mathbf{b}},
\ee
where $c_{a, \mathbf{b}}$ is a constant to be determined by the condition $S_{a, \mathbf{b}}'(\infty) = 0$. By induction over the depth, one 
obtains all the desired derivatives. For instance
\be
S_{a}'(N) = -a\,S_{a+1} + c_{a} = a\,(\zeta_{a+1}-S_{a+1}).
\ee
\ba
S_{a, b}'(N) &=& -a\,S_{a+1, b} + \sum_{n=1}^N \frac{1}{n^a}\,S_b'(n) + c_{a, b} = \\
&=& -a\,S_{a+1, b} -b\,S_{a, b+1}+ b\,S_a\,\zeta_{b+1} + c_{a, b},\nonumber,
\ea
with 
\be
c_{a,b} = a\,S_{a+1, b}(\infty) +b\,S_{a, b+1}(\infty)- b\,\zeta_a\,\zeta_{b+1}.
\ee

\end{document}